\documentclass[11pt,twoside]{article}

\usepackage[utf8]{inputenc} 
\usepackage[T1]{fontenc}
\usepackage{url}
\usepackage{ifthen}
\usepackage{cite}
\usepackage[cmex10]{amsmath} 

\newtheorem{theorem}{Theorem}[section]
\newtheorem{lemma}[theorem]{Lemma}
\newtheorem{corollary}[theorem]{Corollary}

\newtheorem{conjecture}[theorem]{Conjecture}

\newtheorem{definition}[theorem]{Definition}
\newtheorem{example}[theorem]{Example}
\newtheorem{remark}[theorem]{Remark}

\usepackage{amsfonts}
\usepackage{graphicx,color}
\usepackage{amssymb}

\begin{document}
\title{Universal Bounds for Size and Energy of Codes of 
Given Minimum and Maximum Distances}

%
%
%
%

\maketitle

\noindent
\textsc{P. Boyvalenkov} \hfill \texttt{peter@math.bas.bg} \\
{\small Institute for Mathematics and Informatics, BAS, Sofia 1113, Bulgaria} \\[3pt]
{\small and Southwestern University, Blagoevgrad, Bulgaria} \\[3pt]
\textsc{P. Dragnev} \hfill \texttt{dragnevp@pfw.edu} \\
{\small Department of Mathematical Sciences, PFW, Fort Wayne, IN 46805, USA} \\[3pt]
\textsc{D. Hardin} \hfill \texttt{doug.hardin@vanderbilt.edu} \\
{\small Department of Mathematics, Vanderbilt University, Nashville, TN, 37240, USA} \\[3pt]
\textsc{E. Saff} \hfill \texttt{edward.b.saff@vanderbilt.edu} \\
{\small Department of Mathematics, Vanderbilt University, Nashville, TN, 37240, USA} \\[3pt]
\textsc{M. Stoyanova} \hfill \texttt{stoyanova@fmi.uni-sofia.bg} \\
{\small Faculty of Mathematics and Informatics, Sofia University ``St. Kliment Ohridski'', Sofia 1164, Bulgaria}

\begin{abstract}
We employ signed measures that are positive definite up to certain degrees to establish Levenshtein-type upper bounds on the cardinality of codes with given minimum and maximum distances, and universal lower bounds on the potential energy (for absolutely monotone interactions) for codes with given maximum distance and cardinality. The distance distributions of  codes that attain the bounds are found in terms of the parameters of Levenshtein-type quadrature formulas. Necessary and sufficient conditions for the optimality of our bounds are derived. Further, we obtain upper bounds on the energy of codes of fixed minimum and maximum distances and cardinality. 
\end{abstract}


\section{Introduction}
\label{intro}

Let $F_q$ be an alphabet of size $q$. We consider codes (sets)
$C \subset F_q^n =\{ (x_1,\ldots,x_n): x_i \in F_q\}$ 
with the Hamming distance $d(x,y)$ between words $x,y \in F_q^n$. In setting of $F_q^n$ as a polynomial metric space \cite{Lev98} the following change 
of the variable
\[ t=1-\frac{2d}{n} \in T_n:=\left\{t_i=-1+\frac{2i}{n} \, : \, i=0,1,\ldots,n\right\} \]
is very convenient. It brings the distances to "inner"\ products and for $x,y \in F_q^n$ we write 
\[ \langle x,y \rangle = 1-\frac{2d(x,y)}{n}=t_{n-d} \in T_n. \] 

For any code $C \subset F_q^n$ we use
\[s(C):=\max \{\langle x,y \rangle \, : \, x,y \in C, x \neq y \} \in T_n , \]
\[\ell(C):=\min \{\langle x,y \rangle \, : \, x,y \in C, x \neq y \} \in T_n, \]
to denote the counterparts of the minimum and maximum distance of $C$, respectively. 
Denote by
\begin{equation}\label{Cnq_def} C_{n,q}(\ell,s):=\{ C \subset F_q^n | s(C) \le s, \ell(C)\ge \ell \} \end{equation}
the set of codes in $F_q^n$ with pairwise  distances  greater than or equal to the {\it minimum distance} $d:=n(1-s)/2$ and 
less than or equal to the {\it maximum distance} $D:=n(1-\ell)/2$. Let 
\[ \mathcal{A}_q(n,\ell,s):=\max \{|C|: C \in C_{n,q}(\ell,s)\} \]
be the maximum possible cardinality of a code from $C_{n,q}(\ell,s)$. The investigation of the quantities like $\mathcal{A}_q(n,\ell,s)$
is one of the classical problems in the coding theory. 

We are interested also in a minimum energy problem which is somewhat more general but 
turns out to be closely related.

\begin{definition}
Given a (potential) function $h(t):[-1,1] \to [0,+\infty]$ and a code $C \subset F_q^n$, the {\em potential energy} 
(also referred to as {\em $h$-energy}) of $C$ is
\[ E_h(C):=\sum_{x, y \in C, x \neq y} h(\langle x,y \rangle). \]
\end{definition}

While we only need the values of $h$ on the discrete set $T_n$ for computing the $h$-energy, 
we further assume that $h$ is {\em (strictly) absolutely monotone} on the interval [-1,1); that is, $h$ and all its derivatives are defined and (positive) nonnegative on this interval. This approach facilitates our investigation and the explanation of our results. 
We remark that the function $F(z)=h(t)$, where $z=n(1-t)/2$, is  
{\em completely monotone} on $(0,n]$ (i.e.,  $(-1)^k F^{(k)}(z)\ge 0$ for all $z\in (0,n]$) if and only if $h$ is absolutely monotone on $[-1,1]$.

For absolutely monotone potentials $h$ we consider the quantity
\[ \mathcal{E}_h(n,M,\ell):=\min \{E_h(C): C \in C_{n,q}(\ell,1-2/n), |C|=M \}, \]
the smallest possible $h$-energy of a code from $C_{n,q}(\ell,1-2/n)$ with prescribed $M$.


General linear programming bounds for quantities like $\mathcal{A}_q(n,\ell,s)$ and $\mathcal{E}_h(n,M,\ell)$ 
were first introduced by Delsarte \cite{Del} (see \cite{DL,Lev98} and references therein) and Yudin \cite{Y}. 
Linear programming bounds for energies of codes and designs in different spaces (including $F_q^n$) were 
investigated for the first time by Ashikhmin-Barg \cite{AB99}, Ashikhmin-Barg-Litsyn \cite{ABL00} (see also \cite{ABL01,ABL01-ieee}
Energies of codes in $F_q^n$ were considered in 2014 by Cohn and Zhao \cite{CZ14} (see also \cite{CK07}) with a focus on (universally) 
optimal codes and by the authors \cite{BDHSS-DCC} who focused on universal bounds. 

In this paper we use linear programming techniques to derive explicit upper bounds for $\mathcal{A}_q(n,\ell,s)$ and lower bounds for
$\mathcal{E}_h(n,M,\ell)$. Our bounds can be computed for all feasible values of $q$, $n$, $s$, and $\ell$, 
which makes them universal in the sense of Levenshtein \cite{Lev98}. We are not aware of such explicit 
universal bounds in the existing literature (see \cite{HKL06} for a particular case) more than 20 years after the chapter \cite{Lev98} by Levenshtein 
and the paper \cite{DL} by Delsarte and Levenshtein.

There is an intricate interplay between the Levenshtein universal bounds for $\mathcal{A}_q(n,-1,s)$ and 
universal lower bounds on $\mathcal{E}_h(n,M,-1)$ in different polynomial metric 
spaces (see \cite{BDHSS-CA} for Euclidean 
spheres $\mathbb{S}^{n-1}$ and
\cite{BDHSS-DCC} for Hamming spaces $F_q^n$). We further that relationship to corresponding bounds for codes from $C_{n,q}(\ell,1-2/n)$
to derive and investigate simultaneously our cardinality and energy bounds. 

For any real polynomial $f(t)$ we consider its expansion in Krawtchouk  polynomials (see Section~\ref{adjacent-poly}),
\[ f(t)=\sum_{i=0}^n f_i Q_i^{(n,q)}(t) \]
(if the degree of the polynomial $f(t)$ exceeds $n$, then $f(t)$ is taken modulo $\prod_{i=0}^{n}(t-t_i)$) and set 
\[ F_{\geq}:=\{ f(t) \, : \, f_0>0, f_i \geq 0, i=1,2,\ldots,n\}. \]
If $f_i>0$ for $i=0,1,2,\ldots,\deg(f)$, then we write $f(t) \in F_>$. The coefficient $f_0$ 
is of special interest and we call it the {\it zeroth} coefficient of $f(t)$. 

Following Delsarte \cite{Del}, we have 
\begin{equation}\label{MaxCodesLP} 
\mathcal{A}_q(n,\ell,s) \leq \min_{f\in \mathcal{F}_{n,\ell,s}} \frac{f(1)}{f_0},
\end{equation}
where
\[ \mathcal{F}_{n,\ell,s}:=\{ f \in F_{\geq} \, : \, f(t) \leq 0, \,  t \in [\ell,s] \}. \]
Similarly, following Yudin \cite{Y}, we have
\begin{equation}\label{MinEnergyLP} 
\mathcal{E}_h(n,M,\ell) \ge \max_{g\in \mathcal{G}^{(h)}_{n,\ell}} M(Mg_0-g(1)), \end{equation}
where 
\[ \mathcal{G}^{(h)}_{n,\ell}:=\{ g\in F_{\geq} \, : \,  g(t) \leq h(t), \, t \in [\ell,1)\}. \] 
 Therefore, major results in this context crucially depend on proper choice and 
investigation of polynomials that optimize \eqref{MaxCodesLP} or \eqref{MinEnergyLP}. 

The Levenshtein bound (see \cite{Lev92,Lev95,Lev98}) and the energy bound \cite{BDHSS-DCC} work for $\ell=-1$ and, of course, 
depend on the properties of Krawtchouk polynomials and their adjacent polynomials which are orthogonal 
with respect to classical positive measures. The case $\ell>-1$, however, already involves more challenging signed measures. 
In this paper we develop the necessary theory of signed measures to be used in the investigation of the optimization 
problems arising from the right hand sides of \eqref{MaxCodesLP} 
and \eqref{MinEnergyLP}. Then we derive and investigate universal upper bounds for $\mathcal{A}_q(n,\ell,s)$ and lower bounds for
$\mathcal{E}_h(n,M,\ell)$.

The paper is organized as follows. In Sections \ref{adjacent-poly} and \ref{signed-measures} we 
introduce the so-called adjacent polynomials and signed measures. Then we  
establish the positive definiteness of the corresponding measures up to appropriate degrees. Properties of the associated orthogonal 
(and adjacent again) polynomials 
are derived and discussed in Section \ref{Lev-type-poly}, where we define Levenshtein-type polynomials $f_{2k}^{n,\ell,s}(t)$ to be used in \eqref{MaxCodesLP}.
A Levenshtein-type quadrature formula is derived with nodes the roots of $f_{2k}^{n,\ell,s}(t)$ to serve in proofs and properties. 
In Section \ref{Lev-type-bounds} we obtain simultaneously Levenshtein-type upper bounds on $\mathcal{A}_q(n,\ell,s)$ and 
(as in the case $\ell=-1$) the strongly related lower bounds on $\mathcal{E}_h(n,M,\ell)$. An important role in the proof is played by what we call the 
$(k,\ell)$-strengthened Krein condition extending the Levenshtein's strengthened Krein condition. Section \ref{codes-on-bounds} 
is devoted to description of codes which 
would attain our bounds. The distance distributions of such codes are found as functions of corresponding quadrature formulas parameters. 
In Section \ref{test-functions} we prove necessary and sufficient conditions for the optimality of our bounds (in other words, for existence of improving 
polynomials from the sets  $\mathcal{F}_{n,\ell,s}$ and $\mathcal{G}^{(h)}_{n,\ell}$). The optimality (or existence of improvements)
happens only simultaneously for both bounds. A
linear programming refinement of our bounds is
discussed in Section \ref{simplex}, where we provide evidence that in most cases the the nodes of our polynomials serve as the best approximation
for the general linear programming solution. Upper bounds on the energy of the codes from $C_{n,q}(\ell,s)$ (including the case $\ell=-1$) of fixed cardinality $M$ 
are derived in Section \ref{section-uub} providing this way a strip where the energies of all such codes belong. Examples are shown in Section \ref{some-examples} 
where we build a Levenshtein-type system of bounds for a fixed $\ell$. 


\section{Krawtchouk and adjacent polynomials}
\label{adjacent-poly}

For fixed $n$ and $q$, the (normalized) Krawtchouk polynomials are
defined by
\[ Q_i^{(n,q)}(t) :=\frac{1}{r_i} K_i^{(n,q)}(z),  \]
where 
\[ z=\frac{n(1-t)}{2} \] 
is a change of the variable between the set $\{0,1,\ldots,n\}$ of the distances in $F_q^n$ and the set $T_n$,
\[ r_i:=(q-1)^i \binom{n}{i}, \ i=0,1,\ldots,n, \]
are certain dimensions of spaces of functions (see also \eqref{ri-norm} below), and 
\[ K_i^{(n,q)}(z):=\sum_{j=0}^i (-1)^j(q-1)^{i-j} \binom{z}{j} \binom{n-z}{i-j},  \]
$i=0,1,\ldots,n$, are the (usual) Krawtchouk polynomials corresponding to $F_q^n$ (see \cite[Section 2.82]{Sze}). 
In the sequel we will omit the index $(n,q)$ in the notation of Krawtchouk polynomials. 

The polynomials $\{Q_i(t)\}_{i=0}^n$ form a basis of    the space  $\mathcal{P}_n$ of real polynomials of degree at most $n$ and satisfy the following three-term recurrence relation 
\[ (t-a_i) Q_i(t)=b_i Q_{i+1}(t)+c_i Q_{i-1}(t), \]
$i=1,2,\ldots,n-1$, with initial conditions $Q_0(t)=1$ and $Q_1(t)=(qt + q - 2)/(2(q-1))$, where 
\[ a_i=-\frac{(q-2)(n-2i)}{qn}, \]
\[ b_i=\frac{2(q-1)(n-i)}{qn}, \ \ c_i=\frac{2i}{qn}.\]

The measure of orthogonality for the system $\{ Q_i(t) \}_{i=0}^n$ is discrete and given by
\begin{equation} \label{KrawOrtho} \mu_n  := q^{-n}\sum_{i=0}^n r_{n-i} \delta_{t_i},\end{equation}
where $\delta_{t_i} $ is the Dirac-delta measure at $t_i \in T_n$. The form
\[ \langle f,g \rangle=\int f(t) g(t) d\mu_n (t) \]
defines an inner product over the class of polynomials of degree at most $n$. Note that
\begin{equation}
\label{ri-norm}
r_i = \left(\int_{-1}^1 \left(Q_i(t)\right)^2 d \mu_n(t)\right)^{-1}=\|Q_i\|^{-2}.
\end{equation}

We also need $(1,0)$ and $(1,1)$ {\sl adjacent polynomials} as introduced by Levenshtein (cf. \cite[Section 6.2]{Lev98}, see also \cite{Lev92,Lev95}).
Denote
\begin{equation}
\label{kernelT00}
T_i (x,y) := \sum_{j=0}^i r_j Q_j(x)  Q_j(y),
\end{equation}
and define \cite[Eq. (5.65)]{Lev98}
\begin{equation} 
\label{adjacent-10} 
Q_i^{1,0}(t) := \frac{T_i(t,1)}{T_i(1,1)}, \ \ i=0,1,\ldots,n-1.
\end{equation}
Similarly, denote 
\begin{equation}
\label{kernelT10}
T_i^{1,0} (x,y) := \sum_{j=0}^i r_j^{1,0} Q_j^{1,0}(x)  Q_j^{1,0}(y), 
\end{equation}
where 
\[ r_j^{1,0}=\frac{\left(\sum_{u=0}^j r_u\right)^2}{{n-1 \choose j} (q-1)^j}, \ j=0,1,\ldots,n-1, \] 
are the $(1,0)$ counterparts of $r_j$, and define \cite[Eq. (5.68)]{Lev98}
\begin{equation} 
\label{adjacent-11} 
Q_i^{1,1}(t) := \frac{T_i^{1,0}(t,-1)}{T_i^{1,0}(1,-1)}, \ \ i=0,1,\ldots,n-2.
\end{equation}
Note that 
\[ r_j^{1,1}=\frac{\left(\sum_{u=0}^j {n-1 \choose u} (q-1)^u \right)^2}{{n-2 \choose j} (q-1)^j}, \ j=0,1,\ldots,n-2, \]
give the explicit formulas for the $(1,1)$ norm of the polynomials $Q_j^{1,1}(t)$ similatly to \eqref{ri-norm}. 

The corresponding measures of orthogonality of the systems $\{Q_i^{1,0}(t)\}_{i=0}^{n-1}$
and $\{Q_i^{1,1}(t)\}_{i=0}^{n-2}$ are
\begin{equation}
\label{KrawOrtho2} 
c^{1,0}(1-t)d\mu_n (t), \quad c^{1,1}(1-t)(1+t)d\mu_n(t),
\end{equation}
respectively, where 
\[ c^{1,0}=\frac{q}{2(q-1)}, \ \ c^{1,1}=\frac{nq^2}{4(n-1)(q-1)} \]
are normalizing constants (see \cite[Section 6.2]{Lev98}). Of course, the adjacent polynomials 
also satisfy corresponding three-term recurrence relations
\[ (t-a_i^{1,\varepsilon}) Q_i^{1,\varepsilon} (t)=b_i^{1,\varepsilon} Q_{i+1}^{1,\ell} (t)+
c_i^{1,\varepsilon} Q_{i-1}^{1,\varepsilon} (t), \]
where $\varepsilon \in \{0,1\}$, $b_i^{1,\varepsilon}>0$ is the ratio of the leading coefficients of $Q_{i+1}^{1,\varepsilon}(t)$ and $Q_i^{1,\varepsilon}(t)$, $c_i^{1,\varepsilon}=r_{i-1}^{1,\varepsilon}b_{i-1}^{1,\varepsilon}/r_i^{1,\varepsilon}>0$ and
$a_i^{1,\varepsilon}=1-b_i^{1,\varepsilon}-c_i^{1,\varepsilon}$. 

Note also the explicit relations \cite{Lev95}
\[ Q_i^{1,0}(t) = \frac{K_i^{(n-1,q)}(z-1)}{\sum_{j=0}^i r_j}, \]
\[ Q_i^{1,1}(t) = \frac{K_i^{(n-2,q)}(z-1)}{\sum_{j=0}^i \binom{n-1}{j} (q-1)^j}, \]
where $z=n(1-t)/2$ as above, between the $(1,0)$ and $(1,1)$ adjacent polynomials and the
usual Krawtchouk polynomials.

For $\ell \in T_n$ we shall introduce below further adjacent polynomials $Q_i^{1,\ell}(t)$ 
as generalizations of $Q_i^{1,1}(t)$ (note that $\ell=-1$ in $Q_i^{1,\ell}(t)$ gives $Q_i^{1,1}(t)$ 
by the definitions in \cite[Eqn. (5.66)]{Lev98} and \eqref{Pi1l'} below).  
Under certain natural conditions the polynomials $Q_i^{1,\ell}(t)$ are orthogonal with respect to a signed measure 
$d\mu_{n,\ell}(t)$ which is defined and investigated below. 
With the next step, we shall use this new series to construct polynomials $Q_i^{1,\ell,s}(t)$ which
are orthogonal with respect to another signed measure $d\mu_{n,\ell,s}(t)$ again to be defined and 
investigated below. Furthermore, the signed measures $d\mu_{n,\ell}(t)$ and $d\mu_{n,\ell,s}(t)$ 
are strong enough to imply properties which 
are crucial for our constructions. Then our Levenshtein-type polynomials will be constructed 
to be applied in \eqref{MaxCodesLP} and, moreover, as in the case $\ell=-1$ \cite{BDHSS-DCC}, to 
setup polynomials to be applied in \eqref{MinEnergyLP}. In all these constructions and investigations, 
the Christoffel-Darboux formula \cite[Chapter 3.2]{Sze} plays an important role.

These Levenshtein-type polynomials  
can  also be viewed as adjacent polynomials summarized by the following sequence: 
\begin{equation} \label{seq-poly}
Q_i \to Q_i^{1,0} \to Q_i^{1,\ell} \to Q_i^{1,\ell,s},
\end{equation}
where each subsequent family of polynomials can be expressed in terms of the previous family using the Christoffel-Darboux formula (see \eqref{adjacent-10}, \eqref{adjacent-11},
\eqref{Pi1l'}, and \eqref{poly_sub}).

We conclude this section with notations for the zeros of the  polynomials from the sequence \eqref{seq-poly}. Let 
\[ t_{i,1}^{a}< t_{i,2}^{a} < \cdots < t_{i,i}^{a} \]
be the zeros of the polynomial $Q_i^{a} (t)$, $i=0,1,\ldots,$, where the index $a$ stands for the pairs 
$(1,0)$, $(1,1)$, $(1, \ell)$, or the triple $(1,\ell,s)$, respectively. 


\section{Positive definite signed measures}
\label{signed-measures}

Signed measures were first used by Cohn and Kumar in \cite{CK07} in the context of linear programming bounds for energy 
of spherical codes. \smallskip

\begin{definition} 
A signed Borel measure $\mu$ on $\mathbb{R}$ for which all polynomials are integrable
is called {\it positive definite up to degree $m$} if for all real polynomials $p \not\equiv 0$
of degree at most $m$ we have $\int p^2 (t) d \mu(t) > 0$.  For such $\mu$, the bi-linear form
\begin{equation}\label{ipsm}
\langle f, g\rangle_{\mu}:=\int f(t)g(t)\, d\mu(t),
\end{equation}
is an inner product on the space $\mathcal{P}_m$.
\end{definition} \smallskip

Suppose that $k$,  $\ell$, and $s$ are such that the roots of $Q_k^{1,0}(t)$ lie in the open interval $(\ell,s)\subset(-1,1)$; that is, (note $\ell<0$)
\begin{equation}
\label{l-s-range}
-1\le \ell < t_{k,1}^{1,0}<t_{k,k}^{1,0}<s\le 1.
\end{equation}
Then we define the following signed measures on $[-1,1]$ (see \eqref{KrawOrtho} and \eqref{KrawOrtho2})
\begin{eqnarray}
\label{SignedMeasures1}
d\mu_{n,\ell}(t) &:=& c^{1,\ell}(t-\ell)(1-t)d\mu_n(t),\\
\label{SignedMeasures1.5}
 d\mu_{n,s}(t)&:=&c^{1,s}(s-t)(1-t)d \mu_n(t),\\
\label{SignedMeasures2}
d\mu_{n,\ell,s}(t) &:=& c^{1,\ell,s} (t-\ell)(s-t)(1-t)d\mu_n(t).
\end{eqnarray}
The normalizing constants  in \eqref{SignedMeasures1}--\eqref{SignedMeasures2} are given by 
\[ c^{1,\ell} := \frac{nq^2}{2(q-1)(2(n-1)-nq(1+\ell))}, \]
\[c^{1,s} := \frac{n q^2}{2(q-1)[nq(1+s)-2(n-1)]},\]
\[ c^{1,\ell,s}:=\frac{n^2q^3}{2(q-1)[4(n-1)(nqd_1-n-q+2)-n^2q^2d_2 )]}, \] 
where $d_1=(2+\ell+s)/2$ and $d_2=(1+\ell)(1+s)$. We will show below that for $n$, $q$, $\ell$ and $s$ that satisfy \eqref{l-s-range}
the constants $c^{1,\ell}$,  $c^{1,s}$, and $c^{1,\ell,s}$ are all positive.

The following theorem establishes the positive definiteness of the signed measures \eqref{SignedMeasures1}--\eqref{SignedMeasures2} 
up to degrees $k-1$, $k-1$, and $k-2$, respectively, as a consequence of the appropriate location 
\eqref{l-s-range} of $\ell$ and $s$.
\smallskip

\begin{theorem} \label{lem_pos_def} For given positive integers $n \geq 2$ and  $q \geq 2$, let $k$, 
 $\ell$, and $s$  satisfy the inequalities \eqref{l-s-range}.
Then the measures $\mu_{n,\ell}$,  $\mu_{n,s}$ and $ \mu_{n,\ell,s}$ are
positive definite up to degrees $k-1$, $k-1$, and $k-2$, respectively.
\end{theorem} \smallskip

{\it Proof.} Modifying the classical Radau quadrature \cite[Sec. 2.7]{DR} for integration with respect to discrete measures we conclude that
the zeros of the corresponding discrete orthogonal polynomial, the system of $k+1$ nodes
\[ t_{k,1}^{1,0}< t_{k,2}^{1,0}< \dots< t_{k,k}^{1,0}<1 \]
defines a positive (i.e., the weights $w_i$, $i=1,2,\ldots,k+1$, are positive) Radau quadrature with respect to $\mu_n$,
\begin{equation}\label{w_quadrature}
f_0:= \int_{-1}^1 f(t) d\mu_n(t)= w_{k+1} f(1)+\sum_{i=1}^k w_i f(t_{k,i}^{1,0}),
\end{equation}
that is exact for all polynomials of degree at most $2k$.

Using \eqref{w_quadrature} for $f(t)=(t-\ell)(1-t)$, we find that
\[
\left( c^{1,\ell}\right)^{-1}=
 \sum_{i=1}^k w_i (t_{k,i}^{1,0} -\ell)(1-t_{k,i}^{1,0}) > 0.
\]
Similarly, we can show that $c^{1,s}>0$ and $c^{1,\ell,s}>0$.

Next, we apply \eqref{w_quadrature} for $q(t)$, an arbitrary polynomial of degree at most $k-1$, to see that
\begin{eqnarray*}
&& \int_{-1}^1 q^2(t) d\mu_{n,\ell} (t) \\
&=& c^{1,\ell} \int_{-1}^1 q^2(t)(t-\ell)(1-t) d\mu_n(t) \\
&=&  c^{1,\ell} \sum_{i=1}^k w_i q^2(t_{k,i}^{1,0})(t_{k,i}^{1,0} -\ell)(1-t_{k,i}^{1,0}) \geq 0.
\end{eqnarray*}
The equality holds only if $q(t_{k,i}^{1,0} )=0$ for all $i=1,\dots,k$,
which would imply that $q(t) \equiv 0$. Therefore the measure $d\mu_{n,\ell} (t)$ is positive definite up to degree $k-1$. 
That $\mu_{n,s}$ is positive definite up to degree $k-1$ provided $s>t_{k,k}^{1,0}$ follows similarly.

Finally, if $q(t) \not\equiv 0$ has degree at most $k-2$, then we utilize \eqref{w_quadrature} again to see that
\begin{eqnarray*}
&& \int_{-1}^1 q^2(t) d\mu_{n,\ell,s} (t) \\
&=& c^{1,\ell,s} \int_{-1}^1 q^2(t)(t-\ell)(s-t)(1-t) d\mu_n(t) \\
&=&   c^{1,\ell,s} \sum_{i=1}^k w_i q^2(t_{k,i}^{1,0})(t_{k,i}^{1,0} -\ell)(s-t_k^{1,0})(1-t_{k,i}^{1,0})>0.
\end{eqnarray*}
This implies that the measure $d\mu_{n,\ell,s}(t)$ is positive definite up to degree $k-2$, as required.
\hfill $\Box$ \smallskip

Theorem \ref{lem_pos_def} allows us to define orthogonal polynomials
with respect to the corresponding signed measures. This provides essential ingredients for modifying Levenshtein's framework.
\smallskip


\section{Construction of Levenshtein-type polynomials}
\label{Lev-type-poly}

\subsection{Existence and uniqueness of $ Q_j^{1,\ell} (t)$, $j=0,1,\ldots,k$, and $Q_j^{1,\ell,s} (t)$, $j=0,1,\ldots,k-1$}

Some of the basic properties of orthogonal polynomials are no longer valid for series of 
polynomials generated by signed measures. Fortunately, our measures $d\mu_{n,\ell}(t)$ and $d\mu_{n,\ell,s}(t)$ possess the necessary 
properties by Theorem \ref{lem_pos_def}. 
Applying Gram-Schmidt orthogonalization we derive the
existence and uniqueness (for the so-chosen normalizations) of the following two classes
of orthogonal polynomials thus completing the sequence \eqref{seq-poly}. \smallskip

\begin{theorem} \label{cor_ortho} For given positive integers $n \geq 2$, $q \geq 2$, let $k$, 
  $\ell$, and $s$  satisfy the inequalities \eqref{l-s-range}.
The following two classes of orthogonal polynomials are well-defined:
\[ \{ Q_j^{1,\ell} (t) \}_{j=0}^k, \ {\rm w.r.t.} \ d\mu_{n,\ell} (t), \  Q_j^{1,\ell}(1)=1; \]
\[ \{ Q_j^{1,\ell,s} (t) \}_{j=0}^{k-1}, \ {\rm w.r.t.}\ d\mu_{n,\ell,s} (t),  \ Q_j^{1,\ell,s} (1)=1. \]
The polynomials in both classes satisfy a three-term recurrence relation and their zeros interlace.
\end{theorem} \smallskip

For our purposes we shall restrict to values of $\ell$ such that 
\begin{equation}
\label{ell-condition} 
\frac{Q_{k+1}^{1,0}(\ell)}{Q_k^{1,0}(\ell)}<1.
\end{equation}
As shown in the proof of Theorem \ref{rootsofP1l} below 
the condition \eqref{ell-condition} is equivalent to the requirement for the largest 
zero of $Q_k^{1,\ell}(t)$ to be less than $1$.

\subsection{Explicit construction and investigation of the polynomials $Q_j^{1,\ell}(t)$,  $j=0,1,\ldots,k$ } 

The explicit form of the polynomials $Q_j^{1,\ell}(t)$ can be seen as a straightforward generalization 
of \eqref{kernelT00}-\eqref{adjacent-10} by using $\ell$ instead of $-1$.
We utilize the Christoffel-Darboux formula (see, for example \cite[Th. 3.2.2]{Sze}, \cite[Eq. (5.65)]{Lev98})
\begin{equation}
\label{kernelT}
\frac{T_i^{1,0} (x,y)}{r_i ^{1,0}b_i^{1,0}} =  \frac{Q_{i+1}^{1,0}(x)  Q_i^{1,0}(y)-Q_{i+1}^{1,0}(y)  Q_i^{1,0}(x)}{x-y} 
\end{equation}
(when $x=y$ appropriate derivatives are used) in our construction. Moreover, 
similarly to \cite{Lev98}, we use \eqref{kernelT} to prove the
interlacing properties of the zeros of $\{ Q_i^{1,\ell}(t)\}$ with respect to the zeros of $\{Q_i^{1,0}(t)\}$.

In what follows in this and the next sections we assume that 
\begin{equation}
\label{ell-location}
t_{k+1,1}^{1,0}<\ell<t_{k,1}^{1,0}.
\end{equation}

\begin{theorem}
\label{rootsofP1l}
Let $n$, $q$, $k$, and $\ell$ be such that \eqref{ell-condition} and \eqref{ell-location}
 are satisfied. Then 
\begin{equation}
\label{Pi1l'}
Q_i^{1,\ell}(t) = \frac{T_i^{1,0} (t,\ell)}{T_i^{1,0}(1,\ell)}=\eta_i^{1,\ell} t^i + \cdots,\quad i=0,1,\dots, k,
\end{equation}
with $\eta_i^{1,\ell}>0$ and the polynomial $Q_i^{1,\ell}(t)$ has $i$ simple zeros
$t_{i,1}^{1,\ell}<   t_{i,2}^{1,\ell}< \cdots <  t_{i,i}^{1,\ell}$
in the interval $(\ell,1)$. Furthermore, the following interlacing properties 
\begin{equation}\label{Interlacing}\begin{split}
t_{i,j}^{1,\ell} &\in \left(t_{i,j}^{1,0}, t_{i+1,j+1}^{1,0}\right), \ i=1,\dots, k-1, j=1,\dots, i ;\\
t_{k,j}^{1,\ell} &\in \left(t_{k+1,j+1}^{1,0}, t_{k,j+1}^{1,0}\right), \ j=1,\dots,k-1,
\end{split}
\end{equation}
and, finally, $t_{k,k}^{1,\ell} \in \left(t_{k+1,k+1}^{1,0},1\right)$ hold true. 
\end{theorem} \smallskip

{\it Proof.} It follows from \eqref{kernelT} that the kernel $T_i^{1,0} (t,\ell)$
 is orthogonal to any polynomial of degree at most $i-1$
with respect to the measure $\mu_{n,\ell}(t)$. Hence 
\eqref{Pi1l'} follows from the positive definiteness of $d\mu_{n,\ell}(t)$ up to degree $k-1$
and the uniqueness of the Gram-Schmidt orthogonalization process (note also the normalization). The comparison of coefficients 
in \eqref{Pi1l'}  yields $\eta_i^{1,\ell}>0$, $i=0,1,\ldots,k$.

Next, it follows from \eqref{kernelT} and \eqref{Pi1l'} that the solutions of the equation
\begin{equation} \label{FracEq}
\frac{Q_{i+1}^{1,0}(t)}{Q_i^{1,0}(t)} = \frac{Q_{i+1}^{1,0}(\ell)}{Q_i^{1,0}(\ell)}
\end{equation}
are the zeros of $Q_i^{1,\ell}(t)$ and the number $\ell$.  

For every $i<k$ the zeros of $Q_{i+1}^{1,0}(t)$ and $Q_i^{1,0}(t)$ are interlaced and contained in the interval 
$\left[t_{k,1}^{1,0},t_{k,k}^{1,0}\right]$. Since 
${\rm sign}\,Q_i^{1,0}(\ell)=(-1)^i$, we have $Q_{i+1}^{1,0}(\ell)/Q_i^{1,0}(\ell)<0$. The rational function $Q_{i+1}^{1,0}(t)/Q_i^{1,0}(t)$ has simple poles at $t_{i, j}^{1,0}$, $j=1,\dots, i$, and simple zeros at $t_{i+1,j}^{1,0}$, $j=1,\dots, i+1$. Therefore, there is at least one solution $t_{i,j}^{1,\ell}$ of \eqref{FracEq} on each interval $\left(t_{i,j}^{1,0}, t_{i+1,j+1}^{1,0}\right)$, $j=1,\dots,i$, which accounts exactly for the zeros of $Q_i^{1,\ell}(t)$.

When $i=k$ we have $Q_{k+1}^{1,0}(\ell)/Q_k^{1,0}(\ell)>0$. Since $\ell \in \left(t_{k+1,1}^{1,0},t_{k,1}^{1,0}\right)$, we account similarly for 
the first $k-1$ solutions of \eqref{FracEq}, namely $t_{k,j}^{1,\ell} \in \left(t_{k+1,j+1}^{1,0}, t_{k,j+1}^{1,0}\right)$, $j=1,\dots,k-1$,
to establish the interlacing properties \eqref{Interlacing}. For the last zero of $Q_{k}^{1,\ell}(t)$ we use the fact that $Q_{k+1}^{1,0}(t)/Q_k^{1,0}(t)>0$ for $t\in (t_{k+1,k+1}^{1,0},\infty)$. As $\lim_{t\to \infty}Q_{k+1}^{1,0}(t)/Q_k^{1,0}(t) = \infty$, 
we have one more solution $t_{k,k}^{1,\ell}>t_{k+1,k+1}^{1,0}$ of \eqref{FracEq}. Then \eqref{ell-condition} implies that $t_{k,k}^{1,\ell}<1$
because $Q_{k+1}^{1,0}(1)/Q_k^{1,0}(1)=1$. 
\hfill $\Box$ \smallskip


The positive definiteness of the measure $d\mu_{n,\ell}(t)$ implies that
\[r_i^{1,\ell}:=\left( \int_{-1}^1 \left( Q_i^{1,\ell} (t)\right)^2 \,d\mu_{n,\ell} (t) \right)^{-1} >0  \]
 for $i=0,1,\dots,k-1$. The three-term recurrence relation from Theorem \ref{cor_ortho} can be written as
\[ (t-a_i^{1,\ell}) Q_i^{1,\ell} (t)=b_i^{1,\ell} Q_{i+1}^{1,\ell} (t)+c_i^{1,\ell} Q_{i-1}^{1,\ell} (t), \]
$i=1,2, \dots ,k-1$, where 
\[ b_i^{1,\ell}=\frac{\eta_{i+1}^{1,\ell}}{\eta_i^{1,\ell}}>0, \ c_i^{1,\ell}=\frac{r_{i-1}^{1,\ell} b_{i-1}^{1,\ell}}{r_i^{1,\ell}}>0, \ a_i^{1,\ell}=1-b_i^{1,\ell}-c_i^{1,\ell}. \]
The initial conditions are $Q_0^{1,\ell}(t)=1$ and
\[ Q_1^{1,\ell}(t) = \frac{nq(nq\ell+nq-2n+2)t+A}{2B}, \]
where 
\[ A=n^2(q-1)(q\ell+q-2)+n(q\ell+5q-6)-2(q-2), \]
\[ B=n^2(q-2)(q\ell+q-2)+2n(q\ell+4q-3)-4(q-2). \]

Finally in this description we note that by Theorem \ref{cor_ortho} the zeros of the polynomials $Q_j^{1,\ell} (t)$ interlace; i.e.,
\[ t_{j,i}^{1,\ell}<t_{j-1,i}^{1,\ell}<t_{j,i+1}^{1,\ell}, \ i=1,2,\dots, j-1. \] 

We conclude this section with a property of the polynomials $Q_i^{1,\ell}(t)$ which will give a particular answer 
to positive definiteness problems to arise below. \smallskip

\begin{lemma} \label{Pos1} If $\ell$ is as in \eqref{l-s-range}, then $(t-\ell) Q_i^{1,\ell} (t) \in F_>$ 
for every $i=0,1,\ldots,k-1$.  
\end{lemma} \smallskip

{\it Proof.} For every $i=0,1,\ldots,k-1$ it follows from \eqref{kernelT} and \eqref{Pi1l'} that 
\[ (t-\ell)Q_i^{1,\ell} (t)=\frac{1-\ell}{ 1 -q_i } 
\left(Q_{i+1}^{1,0}(t)  -q_i Q_i^{1,0}(t) \right), \]
where $q_i=Q_{i+1}^{1,0}(\ell)/Q_i^{1,0}(\ell)<0$ as in the proof of Theorem \ref{rootsofP1l}. Now 
$Q_j^{1,0} (t) \in F_>$ (this is immediate from the definitions \eqref{kernelT00}-\eqref{adjacent-10}) completes the proof. \hfill
$\Box$ 


\subsection{Construction and investigation of $Q_{k-1}^{1,\ell,s}(t)$ and Levenshtein-type polynomials} 

In this section we perform the next step in our construction. Using the system $\{ Q_i^{1,\ell} (t) \}_{i=0}^k$ 
from the previous section we derive explicitly polynomials $Q_{i}^{1,\ell,s} (t)$, $i=1,2,\ldots,k-1$,
orthogonal with respect to the measure $\mu_{n,\ell,s}(t)$. The last polynomial in this sequence, $Q_{k-1}^{1,\ell,s}(t)$, 
will be the main ingredient in our Levenshtein-type polynomials. 

Consider the Chris\-toffel-Darboux kernel associated with the polynomials $Q_j^{1,\ell}(t)$:
\begin{eqnarray*}
\label{Chr_kernel_q}
R_i^{1,\ell}(x,y) &:=& \sum_{j=0}^i r_j^{1,\ell} Q_j^{1,\ell}(x)  Q_j^{1,\ell} (y) \\
&=& r_i^{1,\ell} b_i^{1,\ell} \frac{Q_{i+1}^{1,\ell} (x)  Q_i^{1,\ell} (y) - Q_{i+1}^{1,\ell} (y)  Q_i^{1,\ell} (x)}{x-y},
\end{eqnarray*}
for $0\leq i \leq k-1$ (when $x=y$ appropriate derivatives are used). Given \eqref{ell-location} and assuming that
\begin{equation}
\label{s-location}
t_{k,k}^{1,0} < s < t_{k,k}^{1,\ell},
\end{equation} 
we define
\begin{equation}
\label{poly_sub}
Q_{i}^{1,\ell,s}(t) := \frac{R_{i}^{1,\ell} (t,s)}{R_{i}^{1,\ell} (1,s)}, \ \ i=0,1,\ldots,k-1.
\end{equation}

We focus on the polynomials $Q_{k-1}^{1,\ell,s} (t)$. Their existence, uniqueness, and the correctness of
the definition \eqref{poly_sub} follow as in Theorem \ref{rootsofP1l}. 
The proof of the next assertion about the zeros of $Q_{k-1}^{1,\ell,s}(t)$ is similar to 
the corresponding part of Theorem \ref{rootsofP1l} but we include it for
convenience of the reader. In addition to \eqref{ell-condition}
we require 
\begin{equation}
\label{s-condition} 
\frac{Q_k^{1,\ell} (s)}{Q_{k-1}^{1,\ell} (s)}>\frac{Q_k^{1,\ell}(\ell)}{Q_{k-1}^{1,\ell}(\ell)}
\end{equation}
in order to get the smallest zero of $Q_{k-1}^{1,\ell,s}(t)$ in the interval $(\ell,t_{k,1}^{1,\ell})$. \smallskip

\begin{theorem}\label{rootsofP1ls}
Let $n$, $q$, $\ell$, $s$, and $k$ be such that \eqref{ell-location}, \eqref{s-location}, \eqref{ell-condition} and \eqref{s-condition} 
are fulfilled. Then the polynomial
$Q_{k-1}^{1,\ell,s} (t)$ has $k-1$ simple zeros $\alpha_1<\alpha_2<\cdots<\alpha_{k-1}$ such that
\[ \alpha_1 \in (\ell,t_{k,1}^{1,\ell}), \ \alpha_{i+1} \in (t_{k-1,i}^{1,\ell},t_{k,i+1}^{1,\ell}), \]
$i=1,2,\ldots,k-2$. In particular, $\ell<\alpha_1$ and $\alpha_{k-1}<s$. 
\end{theorem} \smallskip

{\it Proof.} It follows from the Chris\-toffel-Darboux formula for $R_i^{1,\ell}$ and the definition \eqref{poly_sub} that the solutions of the
equation
\begin{equation}
\label{eq-betas}
\frac{Q_k^{1,\ell} (t)}{Q_{k-1}^{1,\ell} (t)}=\frac{Q_k^{1,\ell} (s)}{Q_{k-1}^{1,\ell} (s)}
\end{equation}
are  the zeros of $Q_{k-1}^{1,\ell,s} (t)$ and the number $s$.

The rational  function $Q_k^{1,\ell} (t)/Q_{k-1}^{1,\ell} (t)$ has $k-1$ simple poles at the
zeros $t_{k-1,i}^{1,\ell}$, $i=1,2,\ldots,k-1$, of $Q_{k-1}^{1,\ell}(t)$, and $k$ zeros 
at the zeros $t_{k,i}^{1,\ell}$, $i=1,2,\ldots,k$, of $Q_{k}^{1,\ell}(t)$. Therefore,
there is a solution of \eqref{eq-betas}; i.e., a zero of $Q_{k-1}^{1,\ell,s} (t)$, in each interval $(t_{k-1,i}^{1,\ell},t_{k,i+1}^{1,\ell})$,
$i=1,2,\ldots,k-2$, which accounts exactly for $k-2$ zeros, say 
$\alpha_2<\alpha_3<\cdots<\alpha_{k-1}$. Moreover, since $Q_k^{1,\ell}(s)/Q_{k-1}^{1,\ell} (s)<0$ 
under the assumptions for $s$, we actually have 
\[ \alpha_{i+1} \in (t_{k-1,i}^{1,\ell},t_{k,i+1}^{1,\ell}), \ i=2,3,\ldots,k-2. \]
Note that $\alpha_{k-1}<t_{k,k-1}^{1,\ell}<t_{k,k}^{1,0}<s$. 

Since the function $Q_k^{1,\ell} (t)/Q_{k-1}^{1,\ell} (t)$ increases
from $-\infty$ to $+\infty$ in the interval $[-\infty,t_{k-1,1}^{1,\ell})$, the inequalities
$0>Q_k^{1,\ell} (s)/Q_{k-1}^{1,\ell} (s)>Q_k^{1,\ell}(\ell)/Q_{k-1}^{1,\ell}(\ell)$ (see \eqref{s-condition}) imply that
the smallest zero $\alpha_1$ of $Q_{k-1}^{1,\ell,s} (t)$ lies in the interval
$(\ell,t_{k,1}^{1,\ell})$.

Finally, using again that $Q_k^{1,\ell}(s)/Q_{k-1}^{1,\ell} (s)<0$ and the fact that 
the function $Q_k^{1,\ell} (t)/Q_{k-1}^{1,\ell} (t)$ strictly
increases from $-\infty$ to 1 for $t \in (t_{k-1,k-1}^{1,\ell},1]$, we have the root $s$ of \eqref{eq-betas} in this interval.
\hfill $\Box$ \smallskip

We can already define the Levenshtein-type polynomial
\begin{equation}\label{f_{2k}}
f_{2k}^{n,\ell,s}(t):= (t-\ell)(t-s)\left( Q_{k-1}^{1,\ell,s} (t)\right)^2
\end{equation}
and proceed with an investigation of its properties.

Denote by $L_i (t)$, $i=0,1,\dots,k+1$, the Lagrange basic polynomials generated by the 
nodes $\alpha_0<\alpha_1<\dots<\alpha_{k-1}<\alpha_k<1$ and define 
\begin{equation}
\label{rhoi-lagrange}
\rho_i:=\int_{-1}^1 L_i(t) d\mu_n(t), \ i=0,1,\dots,k+1.
\end{equation}

The next statement is an analog of one of the main theorems (Theorem 5.39) from \cite{Lev98}.
It involves the zeros of $f_{2k}^{n,\ell,s} (t)$ to form a right end-point Radau quadrature formula with positive weights.
\smallskip

 \begin{theorem} \label{QFtheorem} In the context of Theorem \ref{rootsofP1ls}
let $\alpha_0:=\ell$ and $\alpha_k:=s$. Then the Radau quadrature formula
\begin{equation}
\label{QF}
f_0 = \int_{-1}^1 f(t) d \mu_n(t)   = \rho_{k+1} f(1)+    \sum_{i=0}^{k} \rho_i f(\alpha_i) 
\end{equation}
is exact for all polynomials of degree at most $2k$. Moreover, the
weights $\rho_i$, $i=0,\dots, k$, are positive, and $\rho_{k+1}>0$
provided $(t-\ell)Q_k^{1,\ell}(t) \in F_>$.
\end{theorem} \smallskip

{\it Proof.} It follows from \eqref{rhoi-lagrange} that the formula \eqref{QF} is exact for the Lagrange basis and hence for all polynomials of degree at most $k+1$.
By a polynomial division, any polynomial $f(t)$ of degree at most $2k$ can be written as 
\[ f(t)=q(t)(t-\ell)(t-s)(1-t)Q_{k-1}^{1,\ell,s} (t)+g(t), \]
where $\deg(q)\leq k-2$ and $\deg(g) \leq k+1$.
Then the orthogonality of $Q_{k-1}^{1,\ell,s}(t)$ to all polynomials of degree at
most $k-2$ with respect to the measure $d\mu_{n,\ell,s}(t)$ and the fact
that the right-hand side of \eqref{QF} is the same for $f(t)$ and $g(t)$ 
show the exactness of the quadrature formula \eqref{QF} for $f(t)$. 

We next employ the quadrature formula \eqref{QF} to show the positivity of its weights $\rho_i$.

Using the polynomial $f(t)=(1-t)(t-\ell)\left(Q_{k-1}^{1,\ell,s}(t)\right)^2$ in \eqref{QF} 
we obtain
\[ \rho_k f(s) =\int_{-1}^1 \left( Q_{k-1}^{1,\ell,s} (t)\right)^2 \, d\mu_{n,\ell} (t)>0 \]
because of the positive definiteness (up to degree $k-1$) of the measure $d\mu_{n,\ell}(t)$. 
Now $f(s)>0$ implies $\rho_k>0$. Similarly, with the polynomial 
$f(t)=(1-t)(s-t){\left( {Q_{k-1}^{1,\ell,s}}(t)\right)}^2$ in \eqref{QF} 
and the positive definiteness (up to degree $k-1$) of the measure $d\mu_{n,s}(t)$
(see Theorem \ref{lem_pos_def}) we conclude that $\rho_0 >0$. 

To see that $\rho_i>0$ for $i=1,2,\dots,k-1$, we use the polynomials
$f(t)=(1-t)(t-\ell)(s-t) u_{k-1,i}^2 (t)$, respectively, in \eqref{QF}, where $u_{k-1,i}(t)=Q_{k-1}^{1,\ell,s}(t)/(t-\alpha_i)$. Then 
$\deg(u_{k-1,i})=k-2$ and
the positive definiteness of $d\mu_{n,\ell,s} (t)$ (up to degree $k-2$) yields
\[
\rho_i f(\alpha_i)=\int_{-1}^1 u_{k-1,i}^2 (t) \, d\mu_{n,\ell,s}(t)>0. 
\]
Since $f(\alpha_i)>0$ for this choice of $f$, we derive that $\rho_i>0$.

Finally, we consider the weight $\rho_{k+1}$.
In this case we use $f(t)=f_{2k}^{n,\ell,s}(t)$ in \eqref{QF} and find that
\[ f_0=\rho_{k+1}f(1)=\rho_{k+1}(1-s)(1-\ell). \]
Thus it is enough to see that the zeroth coefficient of $f_{2k}^{n,\ell,s}(t)$ is positive.
We use \eqref{poly_sub} to obtain that $f_0$ is equal to
\begin{eqnarray}
\label{f0}
& \int_{-1}^1 (t-\ell)(s-t)(1-t)Q_{k-1}^{1,\ell,s} (t) \frac{Q_{k-1}^{1,\ell,s}(t)-Q_{k-1}^{1,\ell,s} (1)}{t-1}  d\mu_n(t) \nonumber \\
& \qquad + \int_{-1}^1 (t-\ell)(t-s)Q_{k-1}^{1,\ell,s} (t)  d\mu_n(t) \\
&= \frac{1-s}{1-p_k}   \int_{-1}^1 (t-\ell)\left(  Q_k^{1,\ell} (t)-p_k Q_{k-1}^{1,\ell} (t) \right) \, d\mu_n(t), \nonumber
\end{eqnarray}
where $p_k=Q_k^{1,\ell} (s)/Q_{k-1}^{1,\ell} (s)<0$.
Then, under the assumption $(t-\ell)Q_k^{1,\ell}(t) \in F_>$ and with 
$(t-\ell)Q_{k-1}^{1,\ell}(t) \in F_>$ from Lemma \ref{Pos1}, it follows that the last integrand belongs 
to $F_>$ and in particular its zeroth coefficient is positive. This completes the proof of the theorem. \hfill $\Box$ \smallskip


\begin{remark}
The polynomials $f_{2k}^{n,\ell,s}(t)$ can be also constructed via the system $\{Q_i^{1,s}(t)\}_{i=0}^k$ 
instead of $\{ Q_i^{1,\ell} (t) \}_{i=0}^k$ in the sequence \eqref{seq-poly}. Of course, the resulting 
system $\{Q_i^{n,\ell,s}(t)\}_{i=0}^{k-1}$ is the same. 
\end{remark}


\section{Bounding cardinalities and energies }
\label{Lev-type-bounds}

In the proof of the positive definiteness of his polynomials Levenshtein used (see \cite[(3.88) and (3.92)]{Lev98}) what he called the
strengthened Krein condition
\begin{equation}
\label{SKrein}
(t+1)Q_i^{1,1}(t)Q_j^{1,1}(t) \in F_> 
\end{equation}
for every $i,j \in \{0,1,\ldots,n-3\}$. We need the following modification. \smallskip

\begin{definition}
We say that the polynomials $\{Q_i^{1,\ell}(t)\}_{i=0}^k$ satisfy $(k,\ell)$-strengthened Krein condition if
\begin{equation}
\label{LSK}
(t-\ell)Q_i^{1,\ell}(t)Q_j^{1,\ell}(t) \in F_>
\end{equation}
for every $i,j \in \{0,1,\ldots,k\}$ except possibly for $i=j=k$.
\end{definition} \smallskip

The strengthened Krein condition \eqref{SKrein} holds true in $F_q^n$ for all admissible $i$ and $j$ (see \cite[Lemma 3.25]{Lev98}). 
However, the $(k,\ell)$-strengthened Krein condition \eqref{LSK} is not true for every $\ell$, and for
fixed $\ell$ is true only for relatively small $k$. On the other hand, for fixed $n$, all relevant pairs $(k,\ell)$ are finitely many 
and can be therefore subject to computational checks. Lemma \ref{Pos1} says that the condition is 
satisfied for all pairs $(i,0)$, $i=0,1,\ldots,k-1$.  

The main result in this paper is the following. It includes our
Levenshtein-type upper bound on $\mathcal{A}_q(n,\ell,s)$ as an analog of Theorem 5.42 of \cite{Lev98} 
and its counterpart, a universal lower bound on $\mathcal{E}_h(n,M,\ell)$ as an analog of the 
universal lower bound for $\mathcal{E}_h(n,M,-1)$ from \cite{BDHSS-DCC}. We will use the notation 
\[ S_j=\sum_{i=0}^j r_i=\sum_{i=0}^j (q-1)^i {n \choose i} \]  
for $j \in \{k-1,k,k+1\}$. \smallskip

 \begin{theorem} \label{Lev-ULB}
Let $n$, $q$, $k$, $\ell $, and $s $ 
satisfy the conditions \eqref{ell-condition}, \eqref{ell-location}, \eqref{s-location}, and \eqref{s-condition} and  suppose
the $(k,\ell)$-strengthened Krein condition holds. 
Then the polynomial $f_{2k}^{n,\ell,s}(t)$ belongs to $\mathcal{F}_{n,\ell,s}$ and, therefore,
\begin{equation}
\label{L-like-bound}
 \mathcal{A}_q(n,\ell,s) \leq \frac{f_{2k}^{n,\ell,s}(1)}{f_0}=\frac{1}{\rho_{k+1}}=L_{2k}(n,\ell,s),
\end{equation}
where
\begin{equation*} 
  \small L_{2k}(n,\ell,s)  := 
 \frac{ S_k\left(Q_{k-1}^{1,\ell} (s)-Q_k^{1,\ell} (s)\right)}
{ \frac{r_{k+1}Q_{k+1}(\ell)Q_{k-1}^{1,\ell} (s)}{S_{k+1}\left(Q_{k+1}^{1,0}(\ell)-Q_k^{1,0}(\ell)\right)} -
\frac{r_{k}Q_{k}(\ell)Q_{k}^{1,\ell} (s)}{S_{k-1}\left(Q_{k}^{1,0}(\ell)-Q_{k-1}^{1,0}(\ell)\right)}}.
 \end{equation*}

Furthermore, for fixed $\ell$, for $h$ being an absolutely monotone function on $[-1,1)$, and for $M$ determined
by 
\[ f_{2k}^{n,\ell,s}(1)=Mf_0, \]
the Hermite interpolant\footnote{The notation $g=H(f;h)$ signifies that
$g$ is the Hermite interpolant to the function $h$ at the zeros (taken with their multiplicity) of $f$.}
\[ g_{2k}^{n,\ell,M}(t):=H((t-s)f_{2k}^{n,\ell,s}(t);h) \]
belongs to $\mathcal{G}^{(h)}_{n,\ell}$, and, therefore,
\begin{equation}
\label{ULB-like-bound} \begin{split}
 \mathcal{E}_h(n,M,\ell) & \geq M(Mg_0-g_{2k}^{n,\ell,M}(1)) \\
& =M^2\sum_{i=0}^{k} \rho_i h(\alpha_i).
\end{split}
\end{equation}

The bounds \eqref{L-like-bound} and \eqref{ULB-like-bound} can be attained only simultaneously by codes which have all their 
inner products in the roots of $f_{2k}^{n,\ell,s}(t)$ and which are, in addition, $2k$-designs\footnote{Also known as orthogonal arrays
of strength $2k$.} in $F_q^n$. 
\end{theorem} \smallskip

{\it Proof.} It follows from the definitions \eqref{poly_sub} and \eqref{f_{2k}} that the polynomial $f_{2k}^{n,\ell,s}(t)$ can be written as
\[ c(t-\ell)\left(Q_k^{1,\ell} (t)+c_1 Q_{k-1}^{1,\ell} (t)\right) \sum_{i=0}^{k-1} r_i^{1,\ell} Q_i^{1,\ell} (t)Q_i^{1,\ell}(s), \]
where $r_i^{1,\ell}>0$, $i=0,1,\ldots,k-1$, and the constants $c=(1-s)/(1+c_1)R_{k-1}^{1,\ell}(1,s)$ and $c_1=-Q_k^{1,\ell} (s)/Q_{k-1}^{1,\ell} (s)$ are positive under
the assumptions for $\ell$ and $s$. Since $Q_i^{1,\ell} (s)>0$ for $0 \leq i \leq k-1$, the polynomial $f_{2k}^{n,\ell,s}(t)$
becomes a positive linear combination of polynomials
$(t-\ell)Q_i^{1,\ell}(t)Q_j^{1,\ell}(t)$, where $i \in \{k,k-1\}$ and $j \leq k-1$. 
Therefore $f_{2k}^{n,\ell,s}(t) \in F_>$ because of the $(k,\ell)$-strengthened Krein condition. This and
the obvious $f_{2k}^{n,\ell,s}(t) \leq 0$ for every $t \in [\ell,s]$ implies that $f_{2k}^{n,\ell,s}(t) \in F_{n,\ell,s}$.

To compute the ratio $f_{2k}^{n,\ell,s}(1)/f_0$ we write $f_0$ as in \eqref{f0} and then
use the representation of $(t-\ell)Q_j^{1,\ell} (t)$ by the Christoffel-Darboux formula (see \eqref{kernelT} and \eqref{Pi1l'})
for $j=k-1$ and $k$. The integrand becomes a linear combination of the polynomials $Q_{i}^{1,0}(t)$, $i=k-1,k,k+1$. Since
\[ \int_{-1}^1 Q_{j}^{1,0}(t) d\mu_n(t)=\int_{-1}^1 \frac{T_j(t,1)}{T_j(1,1)} d \mu_n(t)=\frac{1}{S_j}, \]
after simplifications we obtain the explicit form of the bound \eqref{L-like-bound}.

We proceed with the energy bound. 
Denote by $t_1\leq t_2\leq \cdots \leq t_{2k}$ the zeros of $f_{2k}^{n,\ell,s}(t)$ in increasing order and counting their multiplicity; i.e.,
$t_1:=\alpha_0=\ell$, $t_{2i}=t_{2i+1}:=\alpha_i$, $i=1,\dots,k-1$, and $t_{2k}:=\alpha_k=s$.
Then the Newton interpolation formula gives that the polynomial $g_{2k}^{n,\ell,M}(t)$ is a linear combination with nonnegative coefficients
of the constant 1 and the partial products
\[ \prod_{j=1}^{m} (t-t_j), \ \ m=1, 2,\ldots,2k. \]

Since $t_{2i}$, $i=1,\ldots,k$, are the roots of $Q_k^{1,\ell}(t)+\alpha Q_{k-1}^{1,\ell}(t)$
(see \eqref{poly_sub}) it follows from \cite[Theorem 3.1]{CK07} that the partial products
$\prod_{j=1}^{m} (t-t_{2j})$, $m=1,\ldots,k-1$, have positive coefficients when expanded in terms
of the polynomials $Q_{i}^{1,\ell}(t)$, $i=0,1,\ldots,k-1$. Therefore $g_{2k}^{n,\ell,M}(t)$ is a 
linear combination with positive coefficients of
terms $(t-\ell)Q_i^{1,\ell}(t)Q_j^{1,\ell}(t)$, $i,j \in \{0,1,\ldots,k-1\}$, and the
last partial product which is in fact $f_{2k}^{n,\ell,s}(t)$.
Now $g_{2k}^{n,\ell,M}(t) \in F_>$ follows from the validity of the $(k,\ell)$-strengthened Krein condition
and from $f_{2k}^{n,\ell,s}(t) \in F_>$, obtained in the first part of the proof.

Multiple application of the Rolle's theorem implies that $g_{2k}^{n,\ell,M}(t) \leq h(t)$ for every $t \in [\ell,1)$ and therefore
$g_{2k}^{n,\ell,M}(t) \in \mathcal{G}^{(h)}_{n,\ell}$. The explicit form of the bound \eqref{ULB-like-bound} 
via the weights $\rho_i$ and the nodes $\alpha_i$ follows from the quadrature formula \eqref{QF} applied 
for  $g_{2k}^{n,\ell,M}(t)$ and the interpolation conditions $g_{2k}^{n,\ell,M}(\alpha_i)=h(\alpha_i)$, $i=0,1,\ldots,k$. 

There are two kinds of conditions for attaining the general linear programming bounds \eqref{MaxCodesLP} and \eqref{MinEnergyLP} (see, for example,
\cite[Eqs. (32)-(33)]{Lev95} for \eqref{MaxCodesLP}). First, the inner products of distinct points of any attaining code must be 
among the zeros of the polynomial $f(t)$ in \eqref{MaxCodesLP} or the abscissas of the touching/intersection points of the polynomial $g(t)$ and the 
potential function $h(t)$ in \eqref{MinEnergyLP}. 
Second, the complementary slackness conditions $f_iB_i^\prime=0$ (or $g_iB_i^\prime=0$) 
for $i=1,2,\ldots,n$, where $(B_0^\prime,B_1^\prime,\ldots,B_n^\prime)$  
is the MacWilliams transform of the attaining code, have to be satisfied. 

By our construction, the roots of the polynomial $f_{2k}^{n,\ell,s}(t)$ coincide exactly with the
abscissas of the touching/intersection points of the graphs of $g_{2k}^{n,\ell,M}(t)$ and $h(t)$.  
Further, $f_{2k}^{n,\ell,s}(t) \in F_>$ implies (and $g_{2k}^{n,\ell,M}(t) \in F_>$ does as well) 
that $B_i^\prime=0$ for $i=1,2,\ldots,2k$; i.e., any attaining code has to be a $2k$-design. 

Therefore, the bounds \eqref{L-like-bound} and \eqref{ULB-like-bound} can be attained only simultaneously by codes which have all their 
inner products in the roots of $f_{2k}^{n,\ell,s}(t)$ (equivalently, in the abscissas of the touching/intersection points of the graphs of 
$g_{2k}^{n,\ell,M}(t)$ and $h(t)$) and which are, in addition, $2k$-designs in $F_q^n$. 
This completes the proof. \hfill $\Box$ \smallskip

The bound \eqref{L-like-bound} was obtained and investigated for $k=1$  (in our notations) and the corresponding $\ell$ and $s$
by Helleseth, Kl\o{}ve and Levenshtein \cite{HKL06}. In that paper, comparisons with the Levenshtein bound (see \cite{Lev95}) obtained by polynomials 
of degrees 2 and 3, and detailed descriptions of all known codes attaining $L_2(n,\ell,s)$ can be found. We discuss some examples from \cite{HKL06}
in Section \ref{some-examples}. The bound  \eqref{ULB-like-bound} for $k=1$ is given by
\[ E_h(n,M,\ell) \geq M^2(\rho_0h(\ell)+\rho_1h(s), \]
where $\rho_0$ and $\rho_1$ can be computed as shown in Example \ref{k=1dd-example} below. 

For $k>1$, it does not seem customary to consider the bounds \eqref{L-like-bound} and \eqref{ULB-like-bound} for fixed $k$ 
and varying $\ell$ and $s$. Instead, in Sections \ref{test-functions} and \ref{system-of-bounds} we describe them as a system 
of bounds for fixed $\ell>-1$ and varying $k=1,2,\ldots$ and corresponding $s \in I_k^{(\ell)} \subset (t_{k-1,k-1}^{1,\ell},1)$, like the Levenshtein bound is described with fixed $\ell=-1$. 

The optimality of the bounds \eqref{L-like-bound} and \eqref{ULB-like-bound} will be discussed in Section \ref{test-functions}.
\smallskip

\begin{remark} \label{non-int-M}
The above proof of the bound \eqref{ULB-like-bound} does not require $M$ to be integer. In particular, the expression at the right hand side 
of \eqref{ULB-like-bound} is defined for any real $M \in [2,q^n]$. This is customary in certain investigations. 
\end{remark}


\section{Codes attaining the bounds -- conditions, distance distributions}
\label{codes-on-bounds}

Like in the case $\ell=-1$ (see Theorem 5.55 and Remark 5.58 in \cite{Lev98}; also \cite{BD97} for details), codes which attain 
the bounds from Theorem \ref{Lev-ULB} have special combinatorial and geometric properties. 
Also, it is important that the bounds \eqref{L-like-bound} and \eqref{ULB-like-bound} can be attained only 
simultaneously since the conditions of their attaining coincide. 

The conditions $\alpha_i \in T_n$, $i=1,2,\ldots,k-1$, are quite restrictive. For example, they say that all roots of $Q_{k-1}^{1,\ell,s}(t)$
belong to $T_n$. In particular, the roots of $Q_{k-1}^{1,\ell,s}(t)$ must be all rational which is usually a good starting point for deep algebraic 
investigation. However, in this section we focus on the combinatorial meaning of the fact that all inner products of attaining codes 
must belong to the set $\{\alpha_0,\alpha_1,\ldots,\alpha_k\}$. 
\smallskip

\begin{definition}
Let $C \subset F_q^n$ be a code. For fixed $x \in C$ and $t_{n-i} \in T_n$, $i \in \{0,1,\ldots,n\}$, denote by
\[ A_i(x):=|\{ y \in C : \langle x,y \rangle = t_{n-i} \}|, \]
the number of the points of $C$ at distance $i$ from $x$. The system of nonnegative integers 
$(A_i(x): i=0,1,\ldots,n)$ is called {\it distance distribution of $C$ with respect to $x$}. 
\end{definition} \smallskip

It is clear that $A_0(x)=1$ and that $A_i(x) \neq 0$ is possible only for $i \in \{d,d-1,\ldots,D-1,D\}$, 
where $d$ and $D$ are the minimum and maximum distance of $C$, respectively (recall that $s=1-2d/n$ and $\ell=1-2D/n$). 
We show that for codes attaining \eqref{L-like-bound} and \eqref{ULB-like-bound} the whole distance distribution can be computed. 

When dealing with distance distributions, it is convenient to use the following characteristic property of designs in 
polynomial metric spaces (see \cite{Lev95} for Hamming spaces; Equation (1.10) in \cite{FL95} for the 
general case of polynomial metric spaces). A code $C \subset F_q^n$ is a {\em $\tau$-design} if and only if
\begin{equation}
\label{def-des-f0}
\sum_{y \in C} f(\langle x,y \rangle) =f_0|C|
\end{equation}
holds for every $x \in F_q^n$ and every real polynomial $f(t)$ of degree at most $\tau$. 
\smallskip

\begin{theorem}
\label{dd-codes}
If a code $C \subset C_{n,q}(\ell,s)$ attains the bounds \eqref{L-like-bound} and \eqref{ULB-like-bound}, then 
its distance distribution with respect to any point $x \in C$  does not depend on the choice of $x$  
and can be computed from a system of linear equations. Explicitly, we have
\[ A_{\alpha_i}=A_{\alpha_i}(x)=\rho_i |C| \left(=\frac{\rho_i}{\rho_{k+1}} \right), \ i=0,1,\ldots,k. \]
\end{theorem} \smallskip

{\it Proof.} 
Let $C$ be a code that attains the bounds \eqref{L-like-bound} and \eqref{ULB-like-bound}. By Theorem \ref{Lev-ULB} the code $C$ is a $2k$-design, so \eqref{def-des-f0} holds.
Fixing $x \in C$ and grouping the terms with the same $t_{n-i}$ in the left hand side we write \eqref{def-des-f0} as
\begin{equation}
\label{def-des-f0-grouped}
1+ \sum_{i=d}^D A_i(x) f(t_{n-i})=f_0|C|.
\end{equation}

In our case $A_i (x) \neq 0$ is possible only if 
$t_{n-i} \in \{\alpha_0=\ell,\alpha_1,\ldots,\alpha_{k-1},s=\alpha_k\}$ (see Theorem \ref{Lev-ULB}). 
Setting consecutively $f(t)=1,t,t^2,\ldots,t^{k}$ in \eqref{def-des-f0-grouped} yields the Vandermonde-type system
\begin{equation}
\label{system1}
1+\sum_{i=0}^k A_{\alpha_i}(x) \alpha_i^u = b_u |C|, \ u=0,1,\ldots,k,
\end{equation}
where 
\[ b_u=\int_{-1}^1 t^u d \mu_n \] 
is the zeroth coefficient in the Krawtchouk expansion of $t^u$, $u=0,1,\ldots,k$. 

Since the solution of \eqref{system1} is unique, it follows that the distance distribution $\{ A_{\alpha_i}(x): i=0,1,\ldots,k \}$ does not depend 
on the choice of $x \in C$ and can be computed from 
the system \eqref{system1} (so it is uniquely determined by the parameters $n$, $q$, $\ell$, $s$, and $|C|=L_{2k}(n,\ell,s)$). 
Thus we can omit $x$ in the notation of the distance distributions of $C$. 

The combination of \eqref{system1} and the quadrature formula \eqref{QF} gives explicit formulas for the 
distance distributions. Indeed, setting (again!) the polynomials $f(t)=1,t,t^2,\ldots,t^{k}$ in \eqref{QF} produces the system
\begin{equation}
\label{system2}
\rho_{k+1}+\sum_{i=0}^k \rho_i \alpha_i^u =b_u, \ u=0,1,\ldots,k.
\end{equation}
Multiplying all equations of \eqref{system2} by $|C|$ and taking into account that  
\[ \rho_{k+1}|C|=\rho_{k+1}L_{2k}(n,\ell,s)=1 \]
by \eqref{L-like-bound}, we obtain the system \eqref{system1} again but with unknowns $\rho_i|C|$.  
The solutions of both systems must coincide; i.e., $A_{\alpha_i}=\rho_i|C|$, $i=0,1,\ldots,k$, as required. 
\hfill $\Box$ \smallskip

Of course, the formulas from Theorem \ref{dd-codes} have to produce nonnegative integers. Thus they 
in fact yield strong necessary conditions for existence of codes attaining 
\eqref{L-like-bound} and \eqref{ULB-like-bound}. 

The above approach works (to some extent) also for the external (when $x \in F_q^n \setminus C$) distance distributions 
of $C$ with respect to $x$. Then it yields a system of $2k+1$ equations with respect to 
$n$ unknowns $A_1(x),A_2(x),\ldots,A_n(x)$ (note that $A_0(x)=0$). Typically, $n$ is quite larger than $2k+1$
and our system has many solutions. However, the solutions belong to a finite set and it is possible to find them
for subsequent analysis (see, for example, \cite{BK13,BMS17}). Such computations could yield upper bounds on the
covering radius of codes attaining \eqref{L-like-bound} and \eqref{ULB-like-bound}. In fact,
this approach works in general for designs in $F_q^n$ as well (see \cite{BMS17,BK13}).   

We remark also that the computations of distance distributions of attaining codes allows easy derivation 
of the energy of these codes.
\smallskip

\begin{example}
\label{k=1dd-example}
We show how Theorem \ref{dd-codes} works for $k=1$. Assume that $C \subset C_{n,q}(\ell,s)$ attains the
bound $L_2(n,\ell,s)$. Then the system \eqref{system1} (for $k=1$) is solved explicitly as follows.
We have $b_1=(2-q)/q$ and \eqref{system1} becomes 
\[ \left| \begin{array}{l}
A_\ell+A_s = |C|-1 \\ 
\ell A_\ell +s A_s = \frac{(2-q)|C|}{q}-1, \end{array}
\right.\] 
whence we obtain
\[ A_\ell=\frac{q(1+s)(|C|-1)-2|C|}{q(s-\ell)}, \]
\[ A_s=\frac{2|C|-q(1+\ell)(|C|-1)}{q(s-\ell)}. \]
Now $\rho_0=A_\ell/|C|$ and $\rho_s=A_s/|C|$ are computed in turn and the 
energy of $C$ (attaining the bound \eqref{ULB-like-bound}) is given by
\begin{eqnarray*}
E_h(C) &=& |C|^2\left(\rho_0 h(\ell)+\rho_1 h(s)\right) \\
&=& |C|\left(A_\ell h(\ell)+A_s h(s)\right).
\end{eqnarray*} 
\end{example}

\section{On the optimality of the bounds \eqref{L-like-bound} and \eqref{ULB-like-bound}}
\label{test-functions}

In this section we assume that $n$, $q$, $k$, and $\ell \in  \left(t_{k+1,1}^{1,0}, t_{k,1}^{1,0}\right)$ are fixed and 
there exists an interval $I_k^{(\ell)} \subset (t_{k,k}^{1,0},1)$ such that for every $s \in I_k^{(\ell)} $ the bounds
 \eqref{L-like-bound} and \eqref{ULB-like-bound} are optimal in the following sense
\[ L_{2k}(n,\ell,s) = \min \{L_{2j}(n,\ell,s): j \geq 1, s \in I_k^{(\ell)}\}; \]
i.e., the bound $L_{2k}(n,\ell,s)$ is optimal for every $s \in I_k^{(\ell)}$ among all bounds 
$L_{2j}(n,\ell,s)$ (if any with $j \neq k$) . 
Then the image of $I_k^{(\ell)}$ under the function $L_{2k}(n,\ell,s)$ (which is continuous in $s$) 
is a subinterval of $(L_{2k}(n,\ell,t_{k,k}^{1,0}),q^n)$ denoted by $J_{2k}^{(\ell)}$. 

For $s \in I_k^{(\ell)}$ and positive integer $j$, we define 
\[
R_j^{n,\ell}(s):= \frac{1}{L_{2k}(n,\ell,s)}+\sum_{i=0}^{k}\rho_i Q_j(\alpha_i).
\]
Similarly, for $M \in J_{2k}^{(\ell)}$ and positive integer $j$, we define 
\[
S_j^{n,\ell}(M):= \frac{1}{M}+\sum_{i=0}^{k}\rho_i Q_j(\alpha_i), 
\]
where the parameters $(\rho_i,\alpha_i)_{i=0}^k$ come from fixing $s \in I_k^{(\ell)}$ by the equality $M=L_{2k}(n,\ell,s)$. 
Note that the values of $S_j^{n,\ell}(M)$ for integers $M \in J_{2k}^{(\ell)}$
are just particular values of $R_j^{n,\ell}(s)$ for $s \in I_k^{(\ell)}$. On the other hand, it is clear that $M$ could be considered as real variable 
whenever this facilitates an analysis. \smallskip

\begin{remark}
\label{small-test-functions}
The quadrature formula \eqref{QF} applied for $Q_j^{(n,q)}(t)$ with $1 \leq j \leq 2k$ implies 
immediately that $R_j^{n,\ell}(s)=0$ for all $s \in I_k^{(\ell)}$ and that $S_j^{n,\ell}(M)=0$ 
for all $M \in J_{2k}^{(\ell)}$. Thus we sometimes assume (to avoid trivialities) in what follows that $j \geq 2k+1$. 
\end{remark} 
\smallskip

The next theorem gives necessary and sufficient conditions for 
existence of better bounds than \eqref{L-like-bound} and \eqref{ULB-like-bound} (obtained by polynomials
from $\mathcal{F}_{n,\ell,s}$ and  $\mathcal{G}^{(h)}_{n,\ell}$, respectively). Its part (a) is a counterpart of 
Theorem 5.47 from \cite{Lev98} (see also \cite[Theorem 3.1]{BD98}) and its part (b) is a counterpart of Theorem 5.1 
from \cite{BDHSS-DCC}. \smallskip

\begin{theorem} 
\label{test-functions-thm}
{\rm (a)} Given $n$, $q$, $\ell$, $k$, and $s \in I_k^{(\ell)}$, the bound \eqref{L-like-bound} can be improved by a polynomial from $\mathcal{F}_{n,\ell,s}$
if and only if there exists a positive integer $j \geq 2k+1$ such that $R_j^{n,\ell}(s)<0$. In particular,
if  $R_j^{n,\ell}(s) \geq 0$ for every $j \leq m$ then \eqref{L-like-bound} cannot be improved by a polynomial from $\mathcal{F}_{n,\ell,s}$
of degree at most $m$.

{\rm (b)} Given $n$, $q$, $\ell$, $k$, $M \in   J_{2k}^{(\ell)}$, and a strictly absolutely monotone $h$, 
the bound \eqref{ULB-like-bound} can be improved by a polynomial from $\mathcal{G}^{(h)}_{n,\ell}$
if and only if there exists a positive integer $j \geq 2k+1$ such that $S_j^{n,\ell}(M)<0$. In particular,
if  $S_j^{n,\ell}(M) \geq 0$ for every $j \leq m$ then \eqref{ULB-like-bound} cannot be improved by a polynomial from $\mathcal{G}^{(h)}_{n,\ell}$
of degree at most $m$.
\end{theorem}
\smallskip

{\it Proof.}  (a) Assume that $R_j^{n,\ell}(s) \geq 0$ for every positive integer $j$. Let  
$f(t) \in \mathcal{F}_{n,\ell,s}$ and 
\begin{equation}
\label{n0}
f(t)= u(t)+\sum_{j\geq 2k+1}  f_j Q_j(t), \nonumber
\end{equation}
where $u(t)$ has degree at most $2k$ and zeroth coefficient $u_0$. Note that $f(\alpha_i) \leq 0$ for $i=0,1,\ldots,k$, $f_j\ge 0$ for $j\geq 2k+1$, and $f_0=u_0$. 
Applying \eqref{QF} to $u(t)$ and using that $L_{2k}(n,\ell,s)=1/\rho_{k+1}$, we obtain
\begin{eqnarray*}
f_0 &=& u_0= \frac{u(1)}{L_{2k}(n,\ell,s)}+\sum_{i=0}^k \rho_i u(\alpha_i) \\
&=&  \frac{f(1)}{L_{2k}(n,\ell,s)}+\sum_{i=0}^k \rho_i f(\alpha_i)  - \sum_{j \geq 2k+1} f_j R_j^{n,\ell}(s) \\
&\leq&  \frac{f(1)}{L_{2k}(n,\ell,s)}.
\end{eqnarray*}
Therefore, $f(1)/f_0 \geq L_{2k}(n,\ell,s)$; i.e., $f(t)$ does not produce better bound than \eqref{L-like-bound}. 

Let, conversely, $R_j^{n,\ell}(s) < 0$ for some $j \geq 2k+1$. We construct a degree $j$ improving polynomial 
of the form
\[ v(t)=(a(t)+c) f_{2k}^{n,\ell,s}(t)=\sum_{i=0}^j v_i Q_i(t), \] 
where the number $c$ and the polynomial $a(t)$ of degree $j-2k$ 
will be properly chosen.

The polynomial $a(t)$ is immediate -- applying polynomial division we consider the unique 
polynomials $a(t)$ (quotient) and $b(t)$ (remainder) such that
\[ Q_j(t)=a(t) f_{2k}^{n,\ell,s}(t)+b(t), \]
where the remainder $b(t)$ has degree at most $2k-1$. Let 
\[ b(t)=\sum_{i=0}^{2k-1} b_i Q_i(t), \ \ f_{2k}^{n,\ell,s}(t)=\sum_{i=0}^{2k} f_i Q_i(t) \] 
be the Krawtchouk expansions of $b(t)$ and $f_{2k}^{n,\ell,s}(t)$, respectively. 
Then it is easy to see that 
\[ c:=\max \left\{ -\min_{t \in [\ell,s]} a(t), \max_{0 \leq i \leq 2k-1} \frac{b_i}{f_i}, 0 \right\} \]
(recall that $f_i>0$ for every $i=0,1,\ldots,2k$) guaranties that $v(t) \in \mathcal{F}_{n,\ell,s}$. 

Since the polynomial 
\[ v(t)-Q_j(t)=cf_{2k}^{n,\ell,s}(t)-b(t) \]
has degree at most $2k$, its zeroth coefficient $v_0$ (which coincides with the zeroth coefficient of $v(t)$) can be computed from \eqref{QF}. 
We have consecutively
\begin{eqnarray*}
v_0 &=& \frac{cf_{2k}^{n,\ell,s}(1)-b(1)}{L_{2k}(n,\ell,s)}+\sum_{i=0}^k \rho_i \left(cf_{2k}^{n,\ell,s}(\alpha_i)-b(\alpha_i)\right) \\
&=& \frac{v(1)-1}{L_{2k}(n,\ell,s)}-\sum_{i=0}^k \rho_i Q_j(\alpha_i) \\
&=& \frac{v(1)}{L_{2k}(n,\ell,s)}-R_j^{n,\ell}(s)> \frac{v(1)}{L_{2k}(n,\ell,s)}.
\end{eqnarray*}
Therefore $v(1)/v_0 <L_{2k}(n,\ell,s)$; i.e., $v(t)$ improves on $L_{2k}(n,\ell,s)$, which completes the proof of the sufficiency. 

(b) Suppose that $S_j^{n,\ell}(M) \geq 0$ for every positive integer $j$.
Any polynomial $g(t) \in \mathcal{G}^{(h)}_{n,\ell}$ can be written as 
\begin{equation}
\label{n1}
g(t)= u(t)+\sum_{j\geq 2k+1}  g_j Q_j(t) \nonumber
\end{equation}
for some polynomial $u(t)$ of degree at most $2k$ with zeroth coefficient $u_0$. 
We have $g(\alpha_i) \leq h(\alpha_i)$ for $i=0,1,\ldots,k$, $g_j\ge 0$ for every $j\geq 2k+1$, and
$g_0=u_0$. Therefore, using \eqref{QF} for $u(t)$ (recall that $M=L_{2k}(n,\ell,s)$), we consecutively obtain
\begin{eqnarray*}
&& Mg_0- g(1) = Mu_0 - g(1) \\
&=& M\sum_{i=0}^{k} \rho_i u(\alpha_i)-\sum_{j\geq 2k+1} g_j \\
    &=& M\sum_{i=0}^{k} \rho_i \left(g(\alpha_i)-\sum_{j\geq 2k+1} g_j Q_j(\alpha_i)\right)-\sum_{j\geq 2k+1} g_j  \\
    &=& M\sum_{i=0}^{k} \rho_i  g(\alpha_i) - M\sum_{j\geq 2k+1}  g_j \left(\frac{1}{M}+ \sum_{i=0}^{k}\rho_i Q_j(\alpha_i) \right)\\
    &=& M\sum_{i=0}^{k} \rho_i  g(\alpha_i)-M\sum_{j\geq 2k+1} g_jS_j^{n,\ell}(M) \\
    &\le& M\sum_{i=0}^{k} \rho_i h(\alpha_i),
\end{eqnarray*}
where, for the last inequality, we used  $S_j^{n,\ell}(M) \geq 0$ for $j \geq 2k+1$. 
Hence the bound, produced by $g(t)$, does not improve on \eqref{ULB-like-bound}.

Conversely, assume that $h$ is strictly absolutely monotone and suppose
that $S_j^{n,\ell}(M)  <0$ for some positive integer $j \geq 2k+1$. We are going to improve \eqref{ULB-like-bound} by using a polynomial
\[ v(t)=\varepsilon Q_j(t)+a(t)=\sum_{i=0}^j v_i Q_i(t), \]
where the number $\varepsilon >0$ and the polynomial $a(t)$ of degree at most $2k$ will be properly chosen.

Denote 
\[ \tilde{h}(t):= h(t)-\varepsilon Q_j(t) \] 
and select $\varepsilon$ such that
$\tilde{h}^{(i)}(t) \geq 0$ on $[\ell,1]$ for all $i=0,1,\dots,j$. This choice of $\varepsilon$ is possible since the function $h$ is strictly 
absolutely monotone. Since $\tilde{h}^{(i)}(t)=h^{(i)}(t)>0$ for $i>j$ the function $\tilde{h}(t)$ is absolutely monotone. 

Now the polynomial $a(t)$ is chosen to be the Hermite interpolant of the new function $\tilde{h}$ at the nodes $\ell=\alpha_0$ (simply) and 
$\alpha_i$, $i=1,2,\ldots,k$, (doubly) exactly as the original $g_{2k}^{n,\ell,M}(t)$ does.  
Then we can infer as in Theorem \ref{Lev-ULB} that $a(t) \in \mathcal{G}_{n,\ell}^{(\tilde{h})}$ 
implying that $v(t) \in \mathcal{G}^{(h)}_{n,\ell}$. 

It remains to prove that $v(t)$ gives a bound which is better than \eqref{ULB-like-bound} indeed. 
Let 
\[ a(t)=\sum_{i=0}^{2k-1} a_i Q_i(t) \] and note that $v_0=a_0$ and $v(1)=a(1)+\varepsilon$.
We multiply by $\rho_i$ and sum up the interpolation equalities for $a(t)$ to compute
\[ \sum_{i=0}^{k} \rho_i a(\alpha_i)= \sum_{i=0}^{k} \rho_i h(\alpha_i)-\varepsilon \sum_{i=0}^{k} \rho_i Q_j(\alpha_i). \]
Since $$M\sum_{i=0}^{k} \rho_i a(\alpha_i)=Ma_0-a(1)$$ by \eqref{QF} and
$$M\sum_{i=0}^{k} \rho_i Q_j(\alpha_i)=MS_j^{n,\ell}(M)-1 $$
by the definition of the function $S_j^{n,\ell}(M)$, we obtain
\[ Ma_0-a(1)=M\sum_{i=0}^{k} \rho_i h(\alpha_i)+\varepsilon -\varepsilon MS_j^{n,\ell}(M) \]
which yields
\begin{eqnarray*}
Mv_0-v(1) &=& M\sum_{i=0}^{k} \rho_i h(\alpha_i)-\varepsilon MS_j^{n,\ell}(M) \\
&>& M\sum_{i=0}^{k} \rho_i h(\alpha_i).
\end{eqnarray*}
The last inequality means that the polynomial $v(t)$ gives better than \eqref{ULB-like-bound} bound.
\hfill $\Box$ \smallskip

Remark \ref{small-test-functions} and Theorem \ref{test-functions-thm} give the following optimality property of the bounds 
\eqref{L-like-bound} and \eqref{ULB-like-bound}. \smallskip

\begin{corollary}
\label{optimal-1-level}
None of the bounds 
\eqref{L-like-bound} and \eqref{ULB-like-bound} can be improved by using polynomials from $\mathcal{F}_{n,\ell,s}$
and $\mathcal{G}^{(h)}_{n,\ell}$, respectively, of degree at most $2k$. 
\end{corollary} \smallskip

The corresponding optimality results for the case $\ell=-1$ 
were proved for the maximum code problem by Sidelnikov \cite{Sid80} (see also \cite{Lev92}) and 
for the minimum energy problem by the authors \cite{BDHSS-DCC}. 

We provide another formula for the test functions. We use the notations
\[ Q_j(t):=\sum_{i=0}^j a_{j,i} t^i \]
for the coefficients of the Krawtchouk polynomials,
\[ S_u=\frac{1}{L_{2k}(n,\ell,s)}+\sum_{i=0}^k \rho_i \alpha_i^u, \]
and recall that $b_u:=\int_{-1}^1 t^u d \mu_n$ as in the proof of Theorem \ref{dd-codes}. \smallskip

\begin{lemma} \label{aijbi}
With the above notations,
\[ \sum_{i=0}^j a_{j,i} b_i=0 \]
for very positive integer $j$. 
\end{lemma}
\smallskip

{\it Proof.} This is the zeroth coefficient of $Q_j(t)$ which is, of course, equal to 0.
\hfill $\Box$ \smallskip

\begin{lemma} \label{subu}
With the above notations,
\[ b_u=S_u \] 
for every $u=0,1,\ldots,2k$.
\end{lemma}
\smallskip

{\it Proof.} This follows from the quadrature formula \eqref{QF} applied with the polynomial 
$t^u$. \hfill $\Box$ \smallskip

\begin{theorem}
\label{test-formula2}
For every $s \in I_k^{(\ell)}$ and positive integer $j>2k$,  
\[
R_j^{n,\ell}(s)= \sum_{u=2k+1}^{j} a_{j,u} \left(S_u-b_u\right)
\]
and, correspondingly, 
\[ S_j^{n,\ell}(M) =  \sum_{u=2k+1}^{j} a_{j,u} \left(S_u-b_u\right) \] 
with parameters coming from $M=L_{2k}(n,\ell,s) \in  J_{2k}^{(\ell)}$ as in Theorem \ref{Lev-ULB}.
\end{theorem} \smallskip

{\it Proof.} It is enough to prove the formula for $R_j^{n,\ell}(s)$. 
Grouping the powers of $\alpha_i$ in the definition of $R_j^{n,\ell}(s)$ 
and using Lemma \ref{aijbi} yield 
\[ R_j^{n,\ell}(s)= \sum_{u=0}^{j} a_{j,u} \left(S_u-b_u\right). \] 
Now Lemma \ref{subu} implies the required identity. \hfill $\Box$ \smallskip

For fixed $n$, $q$, $\ell$ and $k$, there are only finitely many $s \in T_n \cap I_k^{(\ell}$ 
and finitely many $M \in J_k^{(\ell)}$. Thus a numerical investigation of the signs of the functions 
$R_j^{n,\ell}(s)$ and $S_j^{n,\ell}(M)$ can be performed. 

We conclude this section with a few comments on the possibility for using higher degree polynomials.

Corollary \ref{optimal-1-level} implies (like in the case $\ell=-1$) that 
improvements of the bounds \eqref{L-like-bound} and \eqref{ULB-like-bound} by polynomials are only possible for degrees higher than $2k$. We refer to \eqref{L-like-bound} and \eqref{ULB-like-bound} to as \emph{first level bounds}
and call \emph{second level bounds} any improvement by polynomials from $\mathcal{F}_{n,\ell,s}$ or $\mathcal{G}_{n,\ell}^{(h)}$.

In the proof of Theorem  \ref{test-functions-thm} we, in fact, produced improving polynomials. However, the numerical 
experiments show that these are marginal and are never optimal like the first levels are. A detailed  second level universal bounds based on 
Levenshtein-type quadratures that generalize \eqref{QF} will be developed in a future work (see \cite{BDHSS-s} for the spherical codes case when $\ell=-1$).


\section{On optimal linear programming results}
\label{simplex}

Corollary \ref{optimal-1-level} shows that the bounds \eqref{L-like-bound} and \eqref{ULB-like-bound} 
cannot be improved by using polynomials from $\mathcal{F}_{n,\ell,s}$ and $\mathcal{G}^{(h)}_{n,\ell}$, respectively, 
of degree at most $2k$. However, the requirements $f(t) \leq 0$ (or $g(t) \leq h(t)$, respectively) for every 
$t \in [\ell,s]$ (for every $t \in [\ell,1]$, respectively) are stronger than really necessary. What we need in fact, 
is $f(t) \leq 0$ (or $g(t) \leq h(t)$, respectively) for every $t \in [\ell,s] \cap T_n$ (for every $t \in [\ell,1) \cap T_n$, respectively).
Of course, we always have $\{\ell,s\}=\{t_{n-D},t_{n-d}\} \subset T_n$, but the roots $\alpha_1,\alpha_2,\ldots,\alpha_{k-1}$ 
of the polynomial $f_{2k}^{n,\ell,s}(t)$ are not necessarily in the set $T_n$. This makes a difference for $k>1$ allowing a natural relaxation of our linear programming problems. 

We describe a modification of the polynomials $f_{2k}^{n,\ell,s}(t)$ and $g_{2k}^{n,M,\ell}(t)$ which is going to produce 
better bounds provided the new polynomials are still good for linear programming. 

We replace the double roots $\alpha_1,\alpha_2,\ldots,\alpha_{k-1}$ 
(the touching points, respectively) with their closest neighbours from $T_n$. More precisely, if 
$\alpha_i \in (t_{j-1},t_{j})$ for some integer $j\in \{n-D,n-D+1,\ldots,n-d\}$, then we replace 
the double zero $\alpha_i$ of $f_{2k}^{n,\ell,s}(t)$ by two simple zeros $\gamma_{2i-1} = t_{j-1}$ and $\gamma_{2i}=t_{j}$. 
If $\alpha_i=t_j$, then one can try both $(\gamma_{2i-1},\gamma_{2i}) = (t_{j-1},t_{j})$ and $(t_{j},t_{j+1})$. Finally, setting $\gamma_0:=\ell$ and $\gamma_{2k-1}:= s$, 
we define our refining polynomial for the maximum code problem to be 
\[ f_{\rm ref}(t) :=\prod_{i=0}^{2k-1}(t-\gamma_i)=\sum_{i=0}^{2k} f_i Q_i(t). \]
Then 
\[ g_{\rm ref}(t) :=H(f_{\rm ref}(t)(t-t_{n-d-1});h)= \sum_{i=0}^{2k} g_i Q_i(t) \]
will be our refining polynomial for the minimum energy problem. Note that in $g_{\rm ref}$ we may have intersections at $t_{n-d}$
and at $t_{n-d-1}$ or $t_{n-d+1}$ instead of touching at $s=t_{n-d}$.

The above construction obviously preserves the conditions for $f_{\rm ref}(t)$ and $g_{\rm ref}(t)$ for staying feasible at the points of $T_n$; i.e., 
we still have 
\[ f_{\rm ref}(t) \leq 0, \] 
for every $t \in T_n \cap [\ell,s]$ and 
\[ g_{\rm ref}(t) \leq h(t) \]
for every $t \in T_n \cap [\ell,1)$). 

Therefore, only the positive definiteness of $f_{\rm ref}(t)$ and $g_{\rm ref}(t)$ remains to be investigated. 
We remark that the new polynomials have obviously $f_{2k}>0$ and $g_{2k}>0$. Moreover, it follows from the construction
that for every $i=0,1,\ldots,k$ we have $f_{\rm ref}(\alpha_i) \geq 0$ and $g_{\rm ref}(\alpha_i) \geq h(\alpha_i)$ 
with equality if and only if $\alpha_i \in T_n$. Thus the quadrature formula \eqref{QF} implies that the new polynomials have $f_0>0$ and $g_0>0$.  Furthermore, it also implies that
\[ \frac{f_{2k}^{n,\ell,s}(1)}{f_0}\geq \frac{f_{\rm ref}(1)}{(f_{\rm ref})_0},\ \  g_0-\frac{g_{2k}^{n,\ell,M}(1)}{M} \leq (g_{\rm ref})_0-\frac{g_{\rm ref}(1)}{M},\]
so the bounds \eqref{L-like-bound} and \eqref{ULB-like-bound} are indeed improved as claimed above (provided the new polynomials are still feasible).

Numerical investigation of the remaining feasibility conditions $f_i \geq 0$ ($g_i \geq 0$) for $i=1,2,\ldots,2k-1$ show that they are satisfied in
numerous cases. Moreover, numerics lead us to the following conjecture concerning the relaxation of the linear programing over the discrete subset $[\ell,s]\cap T_n$ or $[\ell,1]\cap T_n$ as introduced above. \smallskip

\begin{conjecture} 
\label{refinement-conj}
For fixed $q \geq 3$, $n$, and $\ell$ there exists a constant $s(q,n,\ell)$ such that whenever
$s\in[-1,s(q,n,\ell)) \cap T_n$ (that is large enough $d/n=(1-s)/2$) the new polynomials $f_{\rm ref}$ and $g_{\rm ref}$
solve the relaxed linear programming in the context above
\end{conjecture} \smallskip

In other words, we conjecture that for most parameters the roots $\alpha_1,\alpha_2,\ldots,\alpha_{k-1}$ 
of the Levenshtein-type polynomial $f_{2k}^{n,\ell,s}(t)$ are the best approximation of the optimal 
nodes for general linear programming. This implies significantly faster computation compared, for example, to the simplex method
(see \cite{BJ,Del-sage}). More detailed investigation in this direction will be considered elsewhere. 


\section{Upper energy bounds for codes with given cardinality and minimum and maximum distance}
\label{section-uub}

For given $q$, $n$, $M$, $s$, and $\ell$, denote by 
\[ C_{n,q}(M,\ell,s):=\{ C \subset F_q^n : |C|=M, s(C)=s, \ell(C)=\ell \} \]
the set of codes in $F_q^n$ of cardinality $M \in [2,q^n]$, minimum distance $d=n(1-s)/2$ and diameter
$D=n(1-\ell)/2$. In this section we derive a universal upper bound on the quantity
\[ \mathcal{U}_h(n,M,\ell,s):=\max \{E_h(C): C \in C_{n,q}(M,\ell,s)\}, \]
where $h$ is absolutely monotone. 

The linear programming problem in this case can be formulated as follows
\begin{equation}
\label{lp-ub-energy}
\mathcal{U}_h(n,M,\ell,s) \leq \min_{p(t) \in \mathcal{Q}_{n,M,\ell,s}^{(h)}} M(p_0M-p(1)), 
\end{equation}
with 
\[ p(t)=\sum_{i=0}^n p_i Q_i(t), \] 
\[ \mathcal{Q}_{n,M,\ell,s}^{(h)}:=\{ p(t) \in F_\leq : p(t) \geq h(t), t \in [\ell,s] \}, \]
where $F_\leq := \{ p(t) : p_i \leq 0, i=1,2,\ldots,n\}$.

We construct polynomials which belong to the set $\mathcal{Q}_{n,M,\ell,s}^{(h)}$ and therefore 
provide upper bounds for $\mathcal{U}_h(n,M,\ell,s)$ by \eqref{lp-ub-energy}. Let $s \in I_k^{(\ell)}$,
$\alpha_0=\ell,\alpha_1,\ldots,\alpha_{k-1},\alpha_k=s$ be the roots of $f_{2k}^{n,\ell,s}(t)$ as above, 
and $\rho_0,\rho_1,\ldots,\rho_k$ are the corresponding weights from \eqref{QF}. 
Note that the parameters of the quadrature \eqref{QF} no longer come with $M$ but with $s$ instead.

We consider 
\begin{equation} 
\label{uub_pol}
p_{2k}^{n,M,\ell,s}(t) := -\lambda f_{2k}^{n,\ell,s}(t)+g_L(t) = \sum_{i=0}^{2k} p_i Q_i(t),
\end{equation}
where $\lambda>0$ is a parameter (to be determined and optimized later) and 
\[ g_L(t):=H(f_{2k}^{n,\ell,s}(t);h(t)) \]
is the Hermite interpolation polynomial to the function $h(t)$ that agrees with $h(t)$ exactly in 
the roots of the Levenshtein-type polynomial $f_{2k}^{n,\ell,s}(t)$ (counted with their multiplicities).

Note that $\deg(g_L) \leq 2k-1$ and therefore $\deg(p_{2k}^{n,M,\ell,s})=2k$. Let 
\[ f_{2k}^{n,\ell,s}(t)= \sum_{i=0}^{2k} f_i Q_i(t), \ \ \ g_L(t)=\sum_{i=0}^{2k-1} g_i Q_i(t) \]
be the Krawtchouk expansions of $f_{2k}^{n,\ell,s}(t)$ and $g_L(t)$, respectively. 

The next theorem is the main result in this section.   \smallskip

\begin{theorem}
\label{uub-thm}
Let $n$, $q$, $k$, $\ell$, and $s$ be such that the conditions of Theorem \ref{Lev-ULB} are fulfilled
and let $C \in C_{n,q}(M,\ell,s)$. Then 
\[ E_h(C) \leq \frac{p_{2k}^{n,M,\ell,s}(1)M\left(M-L_{2k}(n,\ell,s)\right)}{L_{2k}(n,\ell,s)} +M^2\sum_{i=0}^{k} \rho_i h(\alpha_i) \]
for every large enough $\lambda$. In particular, 
\begin{eqnarray} \label{uub}
&& \mathcal{U}_h(n,M,\ell,s) \leq U_{2k}(n,M,\ell,s)  \\
&=& \frac{p_{2k}^{n,M,\ell,s}(1)M\left(M-L_{2k}(n,\ell,s)\right)}{L_{2k}(n,\ell,s)} +M^2\sum_{i=0}^{k} \rho_i h(\alpha_i), \nonumber
\end{eqnarray}
where $\lambda$ is chosen by
\begin{equation}
\label{best-lambda}
\lambda :=\max \left\{ \frac{g_i}{f_i}: 1 \leq i \leq 2k-1 \right\}.
\end{equation}

The bound \eqref{uub} can be attained only by codes which have all their 
inner products in the roots of $f_{2k}^{n,\ell,s}(t)$ and $p_iB_i^\prime=0$ for $i=1,2,\ldots,2k$, 
$(B_0^\prime,B_1^\prime,\ldots,B_n^\prime)$ is the MacWilliams transform of the attaining code. 
\end{theorem} \smallskip

{\it Proof.} Since $p_i =-\lambda f_i+g_i$ and $f_i>0$ for every $i=0,1,\ldots,2k$, it follows that large enough 
$\lambda>0$ will make $p_i \leq 0$ for every $1 \leq i \leq 2k-1$. Adding the obvious $p_{2k}<0$, we 
conclude that $p_{2k}^{n,M,\ell,s}(t) \in F_\leq$. 

Moreover, the absolute monotonicity of $h(t)$ and the interpolation conditions for $g_L(t)$ imply that 
$g_L(t) \geq h(t)$ for $t \in [\ell,s]$. Since $f_{2k}^{n,\ell,s}(t) \leq 0$ for $t \in [\ell,s]$ and 
$p_{2k}^{n,M,\ell,s}(\alpha_i)=g_L(\alpha_i)=h(\alpha_i)$ for every $i=0,1,\ldots,k$, it follows from \eqref{uub_pol} that 
$p(t) \geq h(t)$ for every $t \in [\ell,s]$ (whatever $\lambda>0$ is). Therefore 
$p_{2k}^{n,M,\ell,s}(t) \in  \mathcal{Q}_{n,M,\ell,s}^{(h)}$ for large enough $\lambda$ and it remains to compute the corresponding bound. 

We first note that $L_{2k}(n,\ell,s) \geq M$ follows from the monotonicity of the bound \eqref{L-like-bound}. 
Expressing $p_0$ by the quadrature formula \eqref{QF} and using the interpolation conditions we obtain  
\begin{eqnarray*}
&& p_0M-p_{2k}^{n,M,\ell,s}(1) \\
&=& \left(\frac{M}{L_m(n,s)}-1\right)p_{2k}^{n,M,\ell,s}(1) +M\sum_{i=0}^{k} \rho_i h(\alpha_i),
\end{eqnarray*}
whence we get \eqref{uub} with $\lambda$ still to be optimized. The dependence of the right hand side of \eqref{uub} on the parameter 
$\lambda$ comes from $p_{2k}^{n,M,\ell,s}(1)$ only. 
Since $p_{2k}^{n,M,\ell,s}(1)$ is linear and 
increasing with respect to $\lambda$, the best bound is obtained when $\lambda$ is chosen as in \eqref{best-lambda}; 
i.e., when it is the smallest possible real number which satisfies all conditions $p_i=-\lambda f_i +g_i \leq 0$,
$i=1,2,\ldots,2k-1$, simultaneously. Note that $\lambda>0$ by \eqref{best-lambda} since at least one of the ratios, $g_{2k-1}/f_{2k-1}$,
is positive. 

The description of necessary conditions for attaining codes is similar to that in Theorem \ref{Lev-ULB}.
\hfill $\Box$ \smallskip

\begin{corollary}
\label{strip-energy}
The energy of every code from $C_{n,q}(M,\ell,s)$ belongs to the 
interval $[L,U]$, where 
\[ L:=M^2 \sum_{i=0}^k \rho_i^\prime h(\alpha_i^\prime) \]
(the parameters are determined by $M=L_{2k}(n,\ell,s^\prime)$; i.e., by $M$) and 
\[ U:=\frac{M\left(M-L_{2k}(n,\ell,s)f(1)\right)}{L_{2k}(n,\ell,s)} +M^2\sum_{i=0}^{k} \rho_i h(\alpha_i) \]
(the parameters are determined by $f_{2k}^{n,\ell,s}(t)$; i.e., by $s$), respectively. 
\end{corollary} \smallskip

A modification that adds $\ell$ to the interpolation nodes for $g_L(t)$ (so $\ell$ becomes a double node) works 
in a similar way as in Theorem \ref{uub-thm}. However, it is not difficult to prove that the bound produced is the same. 

It is clear that the refining technique from Section \ref{simplex} can be applied for improving the bounds \eqref{uub} and,
consequently, for shrinking the interval $[L,U]$ from Corollary \ref{strip-energy}. 


\section{Examples}
\label{some-examples}

\subsection{System of bounds for $q=2$ and $\ell=-1+2/n$}
\label{system-of-bounds}

We show as a typical example the mix of 
the Levenshtein bounds  (see \cite[Table 6.3]{Lev98}) and our Levenshtein-type bounds \eqref{L-like-bound} for $\mathcal{A}_2(n,\ell,s)$, where  
\[ \ell=t_1=-1+\frac{2}{n}=\frac{2-n}{n} \]
is fixed (this $\ell$ corresponds to $D=n-1$, the second largest possible diameter). The 
fist four bounds (two Levenshtein bounds and our bounds for $k=1$ and $k=2$) are
explicitely stated. 

For $s \in [\ell,-1/n]$ the first Levenshtein bound
\[ A_2(n,\frac{2-n}{n},s) \leq \frac{s-1}{s} \]
is valid. Our bound \eqref{L-like-bound} for $k=1$
\[ A_2(n,\frac{2-n}{n},s) \leq L_2(n,\frac{2-n}{n},s)=\frac{2(1-s)(n-1)}{1-(n-2)s} \]
is valid for 
\[ s \in \left(t_{1,1}^{1,0},t_{1,1}^{1,\ell}\right]=\left(-\frac{1}{n},\frac{2}{n(n-3)}\right]. \]
Then the next Levenshtein bound
\[ A_2(n,\frac{2-n}{n},s) \leq \frac{(1-s)n(2+(n+1)s)}{1-ns^2} \]
comes for 
\[ s \in \left(t_{1,1}^{1,1},t_{2,2}^{1,0}\right]=\left(0,\frac{\sqrt{n-1}-1}{n}\right]. \]
Our bound \eqref{L-like-bound} for $k=2$ is given by
\begin{eqnarray*} && A_2(n,\frac{2-n}{n},s) \leq L_4(n,\frac{2-n}{n},s) \\
&=& \frac{2(1-s)n(n-1)(9s(n-4)+n-18)^2}{81nAs^3+9Bs^2+Cs+D}, 
\end{eqnarray*}
where 
\begin{eqnarray*}
A &=& n^3-9n^2+25n-18,\\
B &=& 29n^3-207n^2+351n-162, \\
C &=& n^5-38n^4+369n^3-675n^2-648n+972, \\
D &=& -n^4+36n^3-279n^2+81n+162. 
\end{eqnarray*}
It is valid for 
\begin{eqnarray*} 
&&  s \in \left(t_{2,2}^{1,0},t_{2,2}^{1,\ell}\right] \\ &=& \left(\frac{\sqrt{n-1}-1}{n},
\frac{2+\sqrt{n^3-12n^2+41n-26}}{n(n-5)}\right]. 
\end{eqnarray*}

The even weight codes $C \subset F_q^n$, where $n=2m+1$ is odd, attain the bound 
\[ L_{2(m-1)}\left(2m+1,\frac{2-n}{n},\frac{n-4}{n}\right)=2^{2m} \]
and the corresponding bound \eqref{ULB-like-bound}. 

\subsection{Cases $k=1$ and $k=2$}

As mentioned above, the case $k=1$ was considered by Helleseth, Klove and Levenshtein \cite{HKL06}. In our notations, their 
bound (see \cite[Theorem 1]{HKL06}) is 
\[ \mathcal{A}_q(n,\ell,s) \leq \frac{L}{L+4(q-1)(1-n)+2nq(q-1)(s+\ell)}, \]
where $L=nq^2(1-s)(1-\ell)$.

Examples with $\ell=-1$ are covered by the Levenshtein bounds (see, for example, Table 1 in \cite{Lev95}). For $\ell>-1$ and $k=1$, 
we extract the following examples from \cite{HKL06}. 

For $n=6$, $q=2$, $\ell=-1/3$ ($D=4$), $s=1/3$ ($d=2$) an explicit nonlinear code in 
\cite[Example 1]{HKL06} has cardinality $M=16=L_2(6,-1/3,1/3)$. For $n=5$, $q=2$, $\ell=-3/5$ ($D=4$), $s=1/5$ ($d=2$) the 
binary $[5,4,2]$ even weight code in \cite[Example 2]{HKL06} has cardinality $M=16=L_2(5,-3/5,1/5)$. 
For $n=56$, $q=3$, $\ell=-17/28$ ($D=45$), $s=-2/7$ ($d=36$) the Hill (ternary) projective cap (see \cite{Hil73}) 
has cardinality $M=729=L_2(56,-17/28,-2/7)$ and for $n=78$, $q=4$, $\ell=-25/39$ ($D=64$), $s=-17/39$ ($d=56$) 
the Hill (quaternary) projective cap (see \cite{Hil76}) has cardinality $M=729=L_2(56,-17/28,-2/7)$. 
All these codes have $h$-energies which attain \eqref{ULB-like-bound} for the corresponding parameters and for every absolutely monotone $h$. 
In all cases the distance distributions of the attaining codes are easily computed by Theorem \ref{dd-codes} (see Example \ref{k=1dd-example}). 

The ovoids in PG$(3,q)$ (see \cite{CK86}) are codes $C \subset F_q^n$ with parameters 
\[ n=q^2+1, \ d=q^2-q, \ D=q^2, \ |C|=q^4. \]
Thus we have 
\[ s=\frac{1+2q-q^2}{1+q^2}, \ \ell=\frac{1-q^2}{1+q^2} \]
and the bound \eqref{L-like-bound} is attained,
\[ |C|=M=q^4=L_2\left(q^2+1,\frac{1-q^2}{1+q^2}, \frac{1+2q-q^2}{1+q^2}\right). \]
The distance distribution of these codes can be computed by the system \eqref{system1} (for $k=1$; 
as in Example \ref{k=1dd-example})) and is given
by 
\[ A_{\ell}=(q-1)(q^2+1)=n(q-1), \]
\[ A_s=q(q-1)(q^2+1)=nd. \]
Thus, in turn we find
\[ \rho_0=A_{\ell} \rho_2=\frac{n(q-1)}{q^4}, \]
\[ \rho_1=A_s\rho_2=\frac{n(q-1)}{q^3}. \]
Then the energy of $C$ (attaining the bound \eqref{ULB-like-bound} for every absolutely monotone $h$) can be computed 
as
\begin{eqnarray*}
&& E_h(C) = M^2 \left(\rho_0 h(\ell)+\rho_1 h(s)\right) \\
&=& q^4(q^2+1)(q-1) \left(h\left(\frac{1-q^2}{1+q^2}\right)+h\left(\frac{1+2q-q^2}{1+q^2}\right)\right). 
\end{eqnarray*}

Even more interesting example coming from \cite{HKL06} is given by an infinite series of codes 
constructed by Dodunekov, Helleseth, and Zinoviev \cite{DHZ04}. For a prime power $q$ and positive integers $m$ and 
$2 \leq N \leq q^m+1$, the length and the cardinality of the codes from \cite{DHZ04} are given by 
\[ n=\frac{q^m-1}{q-1}N, \ |C|=q^{2m}. \]
Further, for these codes we have  
\[ s=-\frac{(N-2)q^m-2(N-1)q^{m-1}+N}{N(q^m-1)}, \]
\[ \ell=-\frac{q^m-2q^{m-1}+1}{q^m-1}, \]
corresponding to $d=(N-1)q^{m-1}$ and $D=Nq^{m-1}$, respectively. 
For these $\ell$ and $s$, the condition $f_1 \geq 0$ for the polynomial $f_2^{n,\ell,s}(t)$ is satisfied for
$N \geq 1+(q^m-q)/2$ (this corresponds to condition (16) from \cite{HKL06}). For such $N$ and with the above $n$, $\ell$ and $s$ 
we have 
\[ |C|=q^{2m}=L_2(n,\ell,s). \]
The distance distribution of these codes is computed by the system \eqref{system1} for $k=1$ (as in Example \ref{k=1dd-example}). We have 
\[ A_\ell=q^{2m}-1-\frac{q(q^m-1)N}{q-1}, \]
\[ A_s=\frac{q(q^m-1)N}{q-1}. \]
The $h$-energy is given by 
\[ E_h(C)=q^{2m}\left(A_\ell h(\ell) + A_s h(s)\right) \]
and attains the bound \eqref{ULB-like-bound} for every absolutely monotone $h$.

Examples with $k=2$ are already rare. In fact, there are many cases with integer $L_4(n,\ell,s)$ but most of them 
(among the checked) fail to produce integer distance distributions from Theorem{dd-codes}. Two well known attaining codes 
are the projections of the binary Golay codes of lengths 23 and 22. Indeed, the first projection of the binary Golay code has parameters $n=23$, 
$\ell=-9/23$ (i.e., $D=16$), $s=7/23$ (i.e., $d=8$) and 
\[ |C|=2^{11}=L_4(23,-\frac{9}{23},\frac{7}{23}) \]
and the second projection has parameters $n=22$, $\ell=-5/11$ (i.e., $D=16$), $s=3/11$ (i.e., $d=8$) and 
\[ |C|=2^{10}=L_4(22,-\frac{5}{11},\frac{3}{11}). \]

{\sl Acknowledgments.} The first author was partially supported by the National Scientific Program "Information and Communication Technologies for a Single Digital Market in Science, Education and Security (ICTinSES)", financed by the Bulgarian Ministry of Education and Science.  The second author was supported, in part, by the Simons Foundation under CGM \#282207. The work of the third and fourth author was supported, in part, by the U. S. National Science Foundation under grant DMS-1516400. The fifth authors was supported in part by Bulgarian NSF contract DN02/2-2016. A preliminary version of this paper with material from Sections 1-5 appeared in the Proceedings of IEEE ISIT2019.

\end{document}